\documentclass[12pt]{article}
\usepackage{geometry}
\usepackage{graphicx}
\usepackage{epstopdf}
\usepackage{amsfonts}
\usepackage{amssymb}
\usepackage{amsthm}
\usepackage{amsmath}
\usepackage{multirow}
\usepackage{cite}

\def \im {{\rm i}}

\def \e  {{\epsilon}}

\newcommand{\dif}{\mathrm{d}}

\newcommand{\sign}{\mathop{\rm sign}\nolimits}

\newcommand{\konst}{\mathop{\rm const.}\nolimits}
\newcommand{\q}{\varphi}
\newcommand{\ch}{\phi}
\newcommand{\z}{z}
\newcommand{\p}{\vartheta}
\newcommand{\n}{n}
\def \rovno {\!\!\!&=&\!\!\!}

\begin{document}

\title{\bf Interpreting $A$ and $B$-metrics with $\Lambda$ \\
    as gravitational field of a tachyon\\
    in (anti-)de~Sitter universe}

\author{O. Hru\v{s}ka$^1$\thanks{E--mail: {\tt HruskaOndrej(at)seznam.cz}} \ and
J. Podolsk\'y$^1$\thanks{E--mail: {\tt podolsky(at)mbox.troja.mff.cuni.cz}}
\\ \\
\small $^1$Institute of Theoretical Physics, Charles University,\\
\small V Hole\v{s}ovi\v{c}k\'ach 2, 18000 Prague 8, Czech Republic. }

\date{\today}
\maketitle

\begin{abstract}
\noindent We investigate main properties and mutual relations
of the so-called $A$ and $B$-metrics with any value of the
cosmological constant. In particular, we explicitly show that
both the $AII$ and $BI$-metrics are, in fact, the famous
Schwarzschild--(anti-)de~Sitter spacetime (that is the
$AI$-metric) boosted to superluminal speed. Together they form
the complete gravitational field of a tachyon in Minkowski or
(anti-)de~Sitter universe. The boundary separating the $AII$
and $BI$ regions is the Mach--Cherenkov shockwave on which the
curvature is unbounded. We analyze various geometric features
of such spacetimes, we provide their natural physical
interpretation, and we visualize them using convenient
background coordinates and embeddings.
\end{abstract}

\newpage

\section{Introduction}
\label{sc:intro}

In a seminal work~\cite{EhlersKundt62} published in 1962,
J.~Ehlers and W.~Kundt systematically investigated static
vacuum gravitational fields. In particular, they introduced a
classification of all such fields of algebraic type~D, denoting
them as classes $A$ and $B$ (and also $C$, later interpreted
physically as the metric in a static region around uniformly
accelerating black holes \cite{KinnersleyWalker70}).

The $A$-metrics, consisting of three subclasses, were written
in \cite{EhlersKundt62} in the form\footnote{We distinguish
three types of the $A$ and $B$-metrics by roman numbers,
instead of arabic employed in~\cite{EhlersKundt62}.}
\begin{eqnarray}
AI:\quad\dif s^2 \rovno r^2\left(\dif\vartheta^2+\sin^2\vartheta\,\dif\varphi^2\right)+\Big(1-\frac{b}{r}\Big)^{-1}\dif r^2-\Big(1-\frac{b}{r}\Big)\dif t^2\,,\label{AIEK}\\
AII:\quad\dif s^2 \rovno z^2\left(\dif r^2+\sinh^2 r\,\dif\varphi^2\right)+\Big(\frac{b}{z}-1\Big)^{-1}\dif z^2-\Big(\frac{b}{z}-1\Big)\dif t^2\,,\label{AIIEK}\\
AIII:\quad\dif s^2 \rovno z^2\left(\dif r^2+ r^2\,\dif\varphi^2\right)+z\,\dif z^2-\frac{\dif t^2}{z}\,,\label{AIIIEK}
\end{eqnarray}
see also Table~18.2 in
\cite{StephaniKramerMacCallumHoenselaersHerlt03} or Chapter~9
in \cite{GriPod09}. The $AI$-metric is the famous Schwarzschild
solution \cite{Schwarzschild16} describing external vacuum
field of a spherically symmetric static object or black hole.
The $A$-metrics can be generalized to include (for example) a
cosmological constant~$\Lambda$, and can be written in a
unified form
\begin{equation}
\dif s^2= p^2(\epsilon_0-\epsilon_2\,q^2)\,\dif \q^2
 + \frac{p^2}{\epsilon_0-\epsilon_2\,q^2}\,\dif q^2
 -\Big(\epsilon_2+\frac{2n}{p}-\frac{\Lambda}{3}\,p^2\Big)\dif t^2
 +\Big(\epsilon_2+\frac{2n}{p}-\frac{\Lambda}{3}\,p^2\Big)^{-1}\dif p^2\,.
\label{AmetricLambda}
\end{equation}
For ${\epsilon_2=1,-1,0}$, we obtain the $AI$, $AII$ and
$AIII$-metric, respectively, as indicated in
Table~\ref{tbl:EKtoPD}. The $AI$-metric with $\Lambda$ is the
(so called) Schwarzschild--de~Sitter solution, first found by
Kottler \cite{Kottler18}, its standard form is obtained by
${p=r}$, ${q=\cos\p}$, ${n=-m}$, ${\epsilon_0=1}$. The $AII$
and $AIII$-metrics have been described and studied as
``topological black holes'' (see, e.g., \cite{GriPod09} for the
list of references).

The $B$-metrics were introduced in \cite{EhlersKundt62} in the
form
\begin{eqnarray}
BI:\quad\dif s^2 \rovno \Big(1-\frac{b}{r}\Big)^{-1}\dif r^2+\Big(1-\frac{b}{r}\Big)\dif\varphi^2+r^2\left(\dif\vartheta^2-\sin^2\vartheta\,\dif t^2\right),\label{BIEK}\\
BII:\quad\dif s^2 \rovno \Big(\frac{b}{z}-1\Big)^{-1}\dif z^2+\Big(\frac{b}{z}-1\Big)\dif\varphi^2+z^2\left(\dif r^2-\sinh^2 r\,\dif t^2\right),\label{BIIEK}\\
BIII:\quad\dif s^2 \rovno z\,\dif z^2+\frac{\dif\varphi^2}{z}+z^2\left(\dif r^2- r^2\dif t^2\right).\label{BIIIEK}
\end{eqnarray}
Although these metrics also look very simple and have been
known for more than fifty years, they have been paid much less
attention than their counterparts (\ref{AIEK})--(\ref{AIIIEK}).
Analogously to (\ref{AmetricLambda}), it is possible to include
any cosmological constant and write the $B$-metrics in a
unified form
\begin{equation}
\dif s^2= -p^2(\epsilon_0-\epsilon_2\,q^2)\,\dif t^2
 + \frac{p^2}{\epsilon_0-\epsilon_2\,q^2}\,\dif q^2
 +\Big(\epsilon_2+\frac{2n}{p}-\frac{\Lambda}{3}\,p^2\Big)\dif \z^2
 +\Big(\epsilon_2+\frac{2n}{p}-\frac{\Lambda}{3}\,p^2\Big)^{-1}\dif p^2\,.
\label{BmetricLambda}
\end{equation}
For the choice ${\epsilon_2=1,-1,0}$, we obtain the  $BI$,
$BII$ and $BIII$-metric, respectively, as also summarized in
Table~\ref{tbl:EKtoPD}. Moreover, as we demonstrated in
\cite{PodHru17,PodHruGri18}, the
parameter~${\epsilon_0=1,-1,0}$ has \emph{no physical meaning}
because it only changes specific coordinate foliation of the
two-dimensional subspace covered by the $t$ and $q$
coordinates. Without loss of generality we may thus choose any
$\epsilon_0$ to obtain the most suitable form of the metric.

\begin{table}[h]
\begin{center}
\begin{tabular}{| l || c | c | c | c | c | c | c | c | c |}
\hline
 & $\epsilon_2$ & $\epsilon_0$ & $\q$ & $q$ & $t$ & $p$ & $n$ & Equations \\
\hline
\hline
$AI$ & $1$ & $1$ & $\varphi$ & $\cos\vartheta$ & $t$ & $r$ & $-b/2$ & (\ref{AmetricLambda}) $\rightarrow$ (\ref{AIEK}) \\
\hline
$AII$ & $-1$ & $-1$ & $\varphi$ & $\cosh r$ & $t$ & $z$ & $b/2$ & (\ref{AmetricLambda}) $\rightarrow$ (\ref{AIIEK}) \\
\hline
$AIII$ & $0$ & $1$ & $r\sin \varphi$ & $r\cos \varphi$ & $t$ & $z$ & $1/2$ & (\ref{AmetricLambda}) $\rightarrow$ (\ref{AIIIEK}) \\
\hline
\hline
 &  &  & $t$ & $q$ & $z$ & $p$ & $n$ & Equations \\
\hline
\hline
$BI$ & $1$ & $1$ & $t$ & $\cos\vartheta$ & $\varphi$ & $r$ & $-b/2$ & (\ref{BmetricLambda}) $\rightarrow$ (\ref{BIEK}) \\
\hline
$BII$ & $-1$ & $-1$ & $t$ & $\cosh r$ & $\varphi$ & $z$ & $b/2$ & (\ref{BmetricLambda}) $\rightarrow$ (\ref{BIIEK}) \\
\hline
$BIII$ & $0$ & $1$ & $r\sinh t$ & $r\cosh t$ & $\varphi$ & $z$ & $1/2$ & (\ref{BmetricLambda}) $\rightarrow$ (\ref{BIIIEK}) \\
\hline
\end{tabular}
\caption{Transformations between the unified form
(\ref{AmetricLambda}) of $A$-metrics (upper part) or
$B$-metrics (\ref{BmetricLambda}) (lower part) and the original
forms (\ref{AIEK})--(\ref{AIIIEK}) or
(\ref{BIEK})--(\ref{BIIIEK}), respectively, as presented by
Ehlers and Kundt in the case when ${\Lambda=0}$.}
\label{tbl:EKtoPD}
\end{center}
\end{table}

The $A$-metric (\ref{AmetricLambda}) can further be generalized
to include electromagnetic charges, rotation, NUT parameter,
and acceleration. Such generalized black holes are contained in
the large Pleba\'nski--Demia\'nski class \cite{PleDem76} of
\emph{expanding} type~$D$ solutions, see \cite{GriPod06b} for
more details. Interestingly, the $C$-metric is then also
naturally included in this Pleba\'nski--Demia\'nski class.

Moreover, the $B$-metrics can also be considered as a subcase
of the Pleba\'nski--Demia\'nski class of metrics in a
\emph{non-expanding} limit in which the double degenerate null
congruence has zero expansion, shear and twist. Such class has
the form
\begin{equation}
 \dif s^2= \varrho^2\Big(-{\cal Q}\,\dif t^2 +\frac{1}{{\cal Q}}\,\dif q^2 \Big)
  +\frac{{\cal P}}{\varrho^2} \Big(\dif z+2\gamma q\,\dif t \Big)^2
  +\frac{\varrho^2}{{\cal P}}\,\dif p^2 ,
 \label{nonExpPD}
 \end{equation}
 where
 \begin{eqnarray}
 \varrho^2 \rovno  p^2+\gamma^2 \,, \qquad
 {\cal Q}(q) =  \epsilon_0-\epsilon_2\,q^2\,, \label{coeffnonexpPD}\\[3pt]
 {\cal P}(p) \rovno  \big(-(e^2+g^2)-\epsilon_2\gamma^2 +\Lambda\gamma^4\big) +2n\,p +(\epsilon_2 -2\Lambda\gamma^2)\,p^2
 -{\textstyle \frac{1}{3}}\Lambda\,p^4\,, \nonumber
\end{eqnarray}
see Section~16.4 in \cite{GriPod09}. It contains two discrete
geometrical parameters ${\epsilon_0, \epsilon_2=1,-1,0}$, the
cosmological constant $\Lambda$, electric and magnetic charges
$e$ and $g$, mass-like parameter $n$, and additional
parameter~$\gamma$. A thorough investigation of the
corresponding de~Sitter and anti-de~Sitter ``backgrounds'' in
the form (\ref{nonExpPD}) (when ${n,\gamma,e,g=0}$, with
$\Lambda\neq0$) was performed in \cite{PodHru17,Hruska2015}.
The Minkowski ``background'' with ${\Lambda=0}$, and  physical
meaning of all seven independent parameters of
(\ref{nonExpPD}), have been recently clarified in
\cite{PodHruGri18}. Clearly, by setting ${e,g,\gamma=0}$, the
class of metrics (\ref{nonExpPD}), (\ref{coeffnonexpPD})
reduces to the $B$-metrics (\ref{BmetricLambda}).

For both $A$ and $B$-metrics (\ref{AmetricLambda}) and
(\ref{BmetricLambda}), the only nonzero Weyl curvature NP
scalar is
\begin{equation}
\Psi_2=\frac{n}{p^3}\, ,
\label{BmetricPsi2Lambda}
\end{equation}
see \cite{PodHruGri18}. The metrics are thus of algebraic
type~D (or conformally flat when ${n=0}$) and have a curvature
singularity at ${p=0}$. Since the metrics depend on the
fraction ${n/p}$, we may restrict ourselves to ${p>0}$ while
keeping $n$ arbitrary. In order to keep the signature
${(-,+,+,+)}$, we must also constraint the range of $p$ such
that ${{\cal P}(p)>0}$.

As noted already by Ehlers and Kundt in \cite{EhlersKundt62},
the $A$-metrics (\ref{AmetricLambda}) and the $B$-metrics
(\ref{BmetricLambda}) have very similar forms, formally related
by a complex transformation ${\q=\im\,t}$ and ${t=\im\,\z}$,
implying ${\dif\q^2\to-\dif t^2}$ and ${\dif t^2\to-\dif
\z^2}$. However, this seems to be just a ``heuristic trick'', a
specific kind of ``Wick rotation''. It is preferable to avoid
such a formal identification. Instead, following
\cite{Peres70}, in Section~\ref{sc:boost} we will employ a
different approach to relate the $A$ and $B$-metrics. This will
be based on performing a \emph{boost of the source}, a
procedure more acceptable from the physical point of view. In
Section~\ref{sc:ranges} we will investigate the admitted
coordinate ranges and possible extensions of the $B$-metrics.
Subsequently, in Section~\ref{sc:Phys} we will examine main
geometrical properties of the complete gravitational fields of
a tachyonic source, composed of the $AII$ and $BI$-metrics,
with the Mach--Cherenkov shocks. All these results will then be
generalized to any value of the cosmological constant $\Lambda$
in Sections~\ref{sc:boostLambda}--\ref{sc:PhysLambda}.

\section{$AII$ and $BI$-metrics are the Schwarzschild spacetime boosted to infinite speed}
\label{sc:boost}

In 1970, A.~Peres \cite{Peres70} realized that it is possible
to obtain exact gravitational field of a (hypothetical) tachyon
by boosting the classic Schwarzschild source (written in
isotropic coordinates) to superluminal speed. In fact, by this
procedure the Mach--Cherenkov shock-wave is also generated
which separates two distinct regions which are described by the
$AII$ and $BI$-metrics. Such tachyonic counterparts of the
Schwarzschild black hole solution were subsequently studied in
more detail by L.~S.~Schulman~\cite{Schulman71} and
J.~R.~Gott~\cite{Gott74}. Let us first summarize this procedure
by explicitly ``boosting'' the usual form of the Schwarzschild
metric to infinite speed, obtaining thus the gravitational
field of a tachyon. And vice versa: It is possible to
``slow-down'' the tachyonic source of the $AII$ and
$BI$-metrics to zero speed, obtaining thus the usual
Schwarzschild $AI$-metric of a static source. In this sense,
the $AI$, $AII$ and $BI$-metrics are related and, in fact,
``equivalent'' --- they just represent (various regions) of the
graviational field generated by a massive source moving with
all possible velocities, including zero and infinity.

\subsection{Boosting the Schwarzschild ($AI$-)metric to ${v\rightarrow \infty}$}
The Schwarzschild metric in the form (\ref{AmetricLambda}) with
${\epsilon_2=1=\epsilon_0}$, ${\Lambda=0}$, ${n=-m}$
can be written in Cartesian coordinates
\begin{eqnarray}\label{eq:Schw_to_cart}
p=r=\sqrt{X^2+Y^2+Z^2}\,,\qquad q=\cos\p= \frac{Z}{\sqrt{X^2+Y^2+Z^2}}\,,\qquad \tan\q=\frac{Y}{X}\,,\qquad T=t\,,
\end{eqnarray}
as
\begin{eqnarray}\label{eq:Schw_cart}
\dif s^2\rovno -\dif T^2+\dif X^2+\dif Y^2+\dif Z^2\\
&&\!\!\!+\frac{2m}{\sqrt{X^2+Y^2+Z^2}}\Bigg(\dif T^2+\left(1-\frac{2m}{\sqrt{X^2+Y^2+Z^2}}\right)^{-1}\frac{(X\dif X+Y\dif Y+Z\dif Z)^2}{X^2+Y^2+Z^2}\Bigg)\,.\nonumber
\end{eqnarray}
Let us now perform a boost in the $Z$-direction:
\begin{equation}\label{eq:boost_for_Schw}
T=\frac{T'+v\,Z'}{\sqrt{1-v^2}},\qquad Z=\frac{Z'+v\,T'}{\sqrt{1-v^2}}\,.
\end{equation}
Although this boost is only allowed for velocities ${|v|<1}$,
it is interesting to observe that \emph{all terms in the metric} (\ref{eq:Schw_cart})
\emph{that introduce} ${\sqrt{1-v^2}}$ \emph{via $T$ and $Z$ are
quadratic}. It is thus possible to consider the limit
${v\rightarrow \infty}$, resulting in
\begin{equation}\label{eq:swapI}
\lim_{v\rightarrow \infty} T^2=\lim_{v\rightarrow \infty}\frac{(T'+v\,Z')^2}{1-v^2}=-Z'^2\,,\qquad
\lim_{v\rightarrow \infty} Z^2=\lim_{v\rightarrow \infty}\frac{(Z'+v\,T')^2}{1-v^2}=-T'^2\,.
\end{equation}
The ``infinite boost'' thus effectively causes \emph{just a
swap} ${T^2\rightarrow -Z'^2}$ and ${Z^2\rightarrow -T'^2}$, so
that the exact Schwarzschild metric (\ref{eq:Schw_cart})
becomes
\begin{eqnarray}\label{eq:Schw_inf_boost}
\dif s^2\rovno-\dif T'^2+\dif X^2+\dif Y^2+\dif Z'^2\\
&&\!\!\!+\frac{2m}{\sqrt{-T'^2+X^2+Y^2}}\Bigg(\!\!-\dif Z'^2+\left(1-\frac{2m}{\sqrt{-T'^2+X^2+Y^2}}\right)^{-1}\frac{(-T'\dif T'+X\dif X+Y\dif Y)^2}{-T'^2+X^2+Y^2}\Bigg)\,.\nonumber
\end{eqnarray}
In fact, this is the $AII$-metric in the region
${T'^2>Y^2+Z^2}$ for $m$ purely imaginary, and the $BI$-metric
in the complementary region ${T'^2<Y^2+Z^2}$ for $m$ real.

Indeed, the $AII$-metric (\ref{AmetricLambda}) with
${\epsilon_2=-1=\epsilon_0}$, ${\Lambda=0}$, written in
Cartesian coordinates
\begin{equation}\label{eq:AII_transf}
p=\sqrt{T^2-X^2-Y^2}\,,\qquad q=\frac{T}{\sqrt{T^2-X^2-Y^2}}\,,\qquad \tan\q=\frac{Y}{X}\,,\qquad t=Z\,,
\end{equation}
reads
\begin{eqnarray}\label{eq:AII_cart}
\dif s^2\rovno-\dif T^2+\dif X^2+\dif Y^2+\dif Z^2\\
&&\!\!\!+\frac{2n}{\sqrt{T^2-X^2-Y^2}}
\Bigg(\!\!-\dif Z^2+\left(1-\frac{2n}{\sqrt{T^2-X^2-Y^2}}\right)^{-1}\frac{(-T\dif T+X\dif X+Y\dif Y)^2}{-T^2+X^2+Y^2}\Bigg).\nonumber
\end{eqnarray}
This is exactly the boosted Schwarzschild metric
(\ref{eq:Schw_inf_boost}) with the identification ${m=\im
\,n}$, i.e. for purely imaginary mass of the (necessarily
tachyonic) source. Here we consider the principal square root
${\sqrt{-T'^2+X^2+Y^2}=\im\, \sqrt{T'^2-X^2-Y^2}}$. Of course,
the transformation (\ref{eq:AII_transf}) is only valid in the
region ${T^2>X^2+Y^2}$.

Similarly, the $BI$-metric (\ref{BmetricLambda}) with
${\epsilon_2=1}$, ${\epsilon_0=-1}$ (which is a more convenient
coordinate representation than ${\epsilon_0=1}$), ${\Lambda=0}$
is put to Cartesian coordinates in the region
${T^2<X^2+Y^2}$~by
\begin{equation}
p = \sqrt{-T^2+X^2+Y^2}\,,\qquad
q = \frac{T}{\sqrt{-T^2+X^2+Y^2}}\,,\qquad
\tan t = \frac{Y}{X}\,,\qquad
z = Z\,,
\end{equation}
taking the form
\begin{eqnarray}\label{eq:BI_cart}
\dif s^2\rovno-\dif T^2+\dif X^2+\dif Y^2+\dif Z^2\\
&&\!\!\!-\frac{2n}{\sqrt{-T^2+X^2+Y^2}}
\Bigg(-\dif Z^2+\left(1+\frac{2n}{\sqrt{-T^2+X^2+Y^2}}\right)^{-1}\frac{(-T\dif T+X\dif X+Y\dif Y)^2}{-T^2+X^2+Y^2}\Bigg).\nonumber
\end{eqnarray}
Again, this is exactly the boosted Schwarzschild metric
(\ref{eq:Schw_inf_boost}), in this case directly with ${m=-n}$.

\subsection{Slowing the $AII$-metric to ${v\rightarrow 0}$}\label{sbs:BoostAII}
Of course, it is also possible to consider a complementary
procedure. Instead of boosting the Schwarzschild static source
to infinite speed, we can stop the tachyonic source of the
$AII$-metric. This is achieved by performing the
boost\footnote{This is formally the same as
(\ref{eq:boost_for_Schw}) for $v$ replaced by $-1/v$.}
\begin{eqnarray}\label{eq:boost_for_AII}
T=\frac{v\,T'-Z'}{\sqrt{v^2-1}}\,,\qquad Z=\frac{v\,Z'-T'}{\sqrt{v^2-1}}\,.
\end{eqnarray}
of the metric (\ref{eq:AII_cart}). Although
(\ref{eq:boost_for_AII}) is only allowed for ${v>1}$, the
metric is quadratic in $T$ and~$Z$, so that it is possible to
make the limit ${v\rightarrow 0}$, resulting in
\begin{equation}\label{eq:swapII}
\lim_{v\rightarrow 0} T^2 = \lim_{v\rightarrow 0}\frac{(v\,T'-Z')^2}{v^2-1}=-Z'^2\,,\qquad
\lim_{v\rightarrow 0} Z^2 = \lim_{v\rightarrow 0}\frac{(v\,Z'-T')^2}{v^2-1}=-T'^2\,.
\end{equation}
As in the case (\ref{eq:swapI}), the limit ${v\rightarrow 0}$
causes the swap ${T^2\rightarrow -Z'^2}$ and ${Z^2\rightarrow
-T'^2}$. The $AII$-metric (\ref{eq:AII_cart}), valid in
${T^2>X^2+Y^2}$, slowed down to ${v=0}$ is thus
\begin{eqnarray}
\dif s^2\rovno-\dif T'^2+\dif X^2+\dif Y^2+\dif Z'^2\\
&&\!\!\!+\frac{2n}{\sqrt{-X^2-Y^2-Z'^2}}\Bigg(\dif T'^2+\left(1-\frac{2n}{\sqrt{-X^2-Y^2-Z'^2}}\right)^{-1}\frac{(X\dif X+Y\dif Y+Z'\dif Z')^2}{X^2+Y^2+Z'^2}\Bigg).\nonumber
\end{eqnarray}
In the region ${X^2+Y^2+Z'^2>0}$ this is the Schwarzschild
metric (\ref{eq:Schw_cart}) with ${n=\im\,m}$.

\subsection{Slowing the $BI$-metric to ${v\rightarrow 0}$}
Using the boost (\ref{eq:boost_for_AII}) we can similarly stop
the superluminal tachyonic source of the $BI$-metric. Due to
the swap (\ref{eq:swapII}), the metric (\ref{eq:BI_cart}) valid
in the region ${T^2<X^2+Y^2}$ becomes
\begin{eqnarray}
\dif s^2\rovno-\dif T'^2+\dif X^2+\dif Y^2+\dif Z'^2\\
&&\!\!\!-\frac{2n}{\sqrt{X^2+Y^2+Z'^2}}\Bigg(\dif T'^2+\left(1+\frac{2n}{\sqrt{X^2+Y^2+Z'^2}}\right)^{-1}\frac{(X\dif X+Y\dif Y+Z'\dif Z')^2}{X^2+Y^2+Z'^2}\Bigg),\nonumber
\end{eqnarray}
that is the Schwarzschild metric (\ref{eq:Schw_cart}) in the
region ${X^2+Y^2+Z'^2>0}$, simply relabeling ${m=-n}$.

Both the $AII$ and $BI$-metrics can thus be physically
interpreted as (a formal limit of) the Schwarzschild solution
boosted to an infinite speed. And vice versa: the superluminal
sources of the $AII$ and $BI$-metrics can be slowed down and
even stopped, yielding exactly the classic Schwarzschild metric
of a static massive source. In this sense, all these three
exact metrics can be understood as representing the ``same
gravitational field'', the \emph{distinction given only by the
speed of the source} and the \emph{region} of spacetime covered
by the corresponding coordinates.

It can also be seen than in the weak-field limit ${n\to0}$, the
$AII$-metric (\ref{eq:AII_cart}) and the $BI$-metric (\ref{eq:BI_cart})
\emph{together cover the whole
Minkowski spacetime}, except the ``separation boundary''
surface ${T^2=X^2+Y^2}$ between them. Physically, it identifies
the \emph{Mach--Cherenkov shock wave} which will be described
in more detail in Section \ref{sc:Phys}, and extended to
any value of the cosmological constant $\Lambda$ in Section
\ref{sc:PhysLambda}.

\section{Coordinate ranges and extensions of the $B$-metrics}\label{sc:ranges}
It is now important to investigate the admitted coordinate
ranges of the $B$-metrics, and their analytic extensions. We
start with the case ${\Lambda=0}$, more general $B$-metrics
with a cosmological constant~$\Lambda$ will be described in
Section~\ref{sc:rangesLambda}.

\subsection{The $BI$-metric}\label{sbsc:BI}
The $BI$-metric is given by (\ref{BmetricLambda}) with
${\epsilon_2=1}$, ${\Lambda=0}$, that is
\begin{equation}\label{eq:BImetric}
\dif s^2= -p^2(\epsilon_0-q^2)\,\dif t^2
 + \frac{p^2}{\epsilon_0-q^2}\,\dif q^2
 +\Big(1+\frac{2\n}{p}\Big)\dif \z^2
 +\Big(1+\frac{2\n}{p}\Big)^{-1}\dif p^2\,.
\end{equation}
Here ${\z\in\mathbb{R}}$, while the range of ${t,q}$ depends on
$\epsilon_0$. Ehlers and Kundt in \cite{EhlersKundt62}
considered the case ${\epsilon_0=1}$, see
Table~\ref{tbl:EKtoPD}. Due to (\ref{BmetricPsi2Lambda}), the
range of $p$ in (\ref{eq:BImetric}) depends on the sign of
$\n$. For ${\n>0}$ it is ${p\in(0,\infty)}$, with a curvature
singularity at ${p=0}$, while for ${\n<0}$ it is
${p\in(2|n|,\infty)}$. Ehlers and Kundt suggested a possible
analytic extension of the $BI$-metric with ${\n<0}$ across
${p=2|n|}$ in the $p$-coordinate. For the choice
${\epsilon_0=-1}$, this is achieved by the
transformation\footnote{The original Ehlers and Kundt
transformation for ${\epsilon_0=1}$ is
${q=\cosh\tau\,\sin\q}$, ${\tanh t=\tanh\tau/\cos\q}$.}
\begin{eqnarray}\label{eq:BIextension-1}
p=2|\n| /(1-\rho^2)\,,\qquad q=\sinh\tau\,,\qquad \z=2\n\,\zeta\,,\qquad t=\q\,,
\end{eqnarray}
which puts the metric (\ref{eq:BImetric}) into the form
\begin{equation}\label{eq:BIextensionmetric}
\dif s^2=4\n^2\Big[\big(1-\rho^2\big)^{-2}\big(-\dif\tau^2+\cosh^2\tau\,\dif\q^2\big)
+\rho^2\dif \zeta^2
+4\big(1-\rho^2\big)^{-4}\dif\rho^2\Big]\,,
\end{equation}
see metric (2--3.47) in~\cite{EhlersKundt62}. Although the term
${\rho^2\dif \zeta^2}$ vanishes at ${\rho=0}$, there is
\emph{no singularity}. The coordinates
(\ref{eq:BIextensionmetric}) of the $BI$-metric thus better
illustrate the behaviour of the spacetime. At ${\rho=0}$,
corresponding to $p=2|\n|$, the curvature
(\ref{BmetricPsi2Lambda}) reaches its \emph{maximal but finite
value} (in fact, it is impossible to reach the curvature
singularity located at ${p=0}$). It is now straightforward to
extend the range of $\rho$ to \emph{negative values}, so that
$\rho\in(-1,1)$. With ${\rho\to \pm 1}$ the spacetime becomes
asymptotically flat (since ${p\to\infty}$). In this aspect, the
global structure of the $BI$-metric resembles the famous
Einstein--Rosen bridge (wormhole) constructed from the
Schwarzschild ($AI$-)solution between two asymptotically flat
universes, one for ${\rho>0}$ and the second for ${\rho<0}$.
The ``neck of the bridge'' is located at
${\rho=0}$.\footnote{In~\cite{EinsteinRosen35}, the extension
of the Schwarzchild solution was obtained by ${p\equiv
r=\rho^2+2m}$.}

\subsection{The $BII$-metric}\label{sbsc:BII}
Analogously, it is possible to extend the $BII$-metrics. In
particular, the metric (\ref{BmetricLambda}) with
${\Lambda=0}$, ${\epsilon_2=-1}$ and ${\epsilon_0=1}$ (to avoid
superficial coordinate singularities in $q$) is
\begin{equation}\label{eq:BIImetric}
\dif s^2= -p^2(1+q^2)\,\dif t^2
 + \frac{p^2}{1+q^2}\,\dif q^2
 +\Big(\frac{2n}{p}-1\Big)\dif \z^2
 +\Big(\frac{2n}{p}-1\Big)^{-1}\dif p^2\,.
\end{equation}
The allowed ranges of coordinates are ${t,q,\z\in\mathbb{R}}$,
${p\in(0,2n)}$. Correct signature requires ${n>0}$, and the
metric (\ref{eq:BIImetric}) \emph{does not admit flat
(Minkowski) limit} given by ${n \to 0}$. Similarly to
(\ref{eq:BIextension-1}) we may apply the transformation
\begin{eqnarray}\label{eq:BIIextension}
p=2\n /(1+\rho^2)\,,\qquad q=\sinh z\,,\qquad \z=2\n\,\zeta\,,
\end{eqnarray}
so that the $BII$-metric (\ref{eq:BIImetric}) becomes
\begin{equation}\label{eq:BIIextensionmetric}
\dif s^2=4\n^2\Big[\big(1+\rho^2\big)^{-2}\big(\dif z^2-\cosh^2z\,\dif t^2\big)
+\rho^2\dif \zeta^2
+4\big(1+\rho^2\big)^{-4}\dif\rho^2\Big]\,,
\end{equation}
see metric (2--3.48) in~\cite{EhlersKundt62}. Again, there is
no singularity at ${\rho=0}$ corresponding to
${p=2n}$, while the curvature singularity at ${p=0}$
corresponds to ${\rho=\infty}$. Static analytic extension of
the $BII$-metric is thus easily obtained by considering the
full range ${\rho\in\mathbb{R}}$ in
(\ref{eq:BIIextensionmetric}).

\subsection{The $BIII$-metric}\label{sbsc:BIII}
This is obtained from (\ref{BmetricLambda}) by setting
${\epsilon_2=0}$, ${\epsilon_0=1}$, $\Lambda=0$,
\begin{equation}\label{eq:BIIImetric}
\dif s^2= -p^2\,\dif t^2
 + p^2\,\dif q^2
 +\frac{2n}{p}\,\dif \z^2
 +\frac{p}{2n}\,\dif p^2\,,
\end{equation}
where $t,q,\z\in\mathbb{R}$, ${p\in(0,\infty)}$. Necessarily
${n>0}$, and the metric again does not have the Minkowski limit
since ${n = 0}$ is prohibited. As argued in
\cite{GriPod09,PodHruGri18}, it is a special Levi-Civita
solution.

\section{Global structure and physical interpretation}
\label{sc:Phys}

After establishing that both the $AII$-metric and the
$BI$-metric represent specific parts of gravitational field
generated by a superluminal source (tachyon moving along a
spacelike trajectory), it is now necessary to analyze the
global structure of such spacetimes and their relation. In
particular, we must describe the way in which the $AII$-metric
\begin{eqnarray}
\label{AIITach}
\dif s^2=\sigma^2(\dif \p^2+\sinh^2\p\,\dif\q^2)
+\Big(1-\frac{2M}{\sigma}\Big)\dif \z^2
-\Big(1-\frac{2M}{\sigma}\Big)^{-1}\dif \sigma^2\,,
\end{eqnarray}
(which is actually the $AII$-metric (\ref{AmetricLambda}) with
${\epsilon_2=-1=\epsilon_0}$, ${n=M>0}$, ${\Lambda=0}$, using
${p=\sigma>0}$, ${q=\cosh\p}$ and ${t=\z}$) is \emph{combined}
with the $BI$-metric
\begin{eqnarray}\label{BITach}
\dif s^2=p^2(-\dif \tau^2+\cosh^2\tau\,\dif\q^2)+\Big(1-\frac{2M}{p}\Big)\dif \z^2
+\Big(1-\frac{2M}{p}\Big)^{-1}\dif p^2\,,
\end{eqnarray}
(which is the $BI$-metric (\ref{BmetricLambda}) with
${\epsilon_2=1}$, ${\epsilon_0=-1}$, ${n=-M}$ , ${\Lambda=0}$,
applying ${q=\sinh\tau}$, ${t=\q}$).

\subsection{Weak-field limit} \label{ssc:weakfield}
\begin{figure}[t]
\begin{center}
\includegraphics[width=170mm]{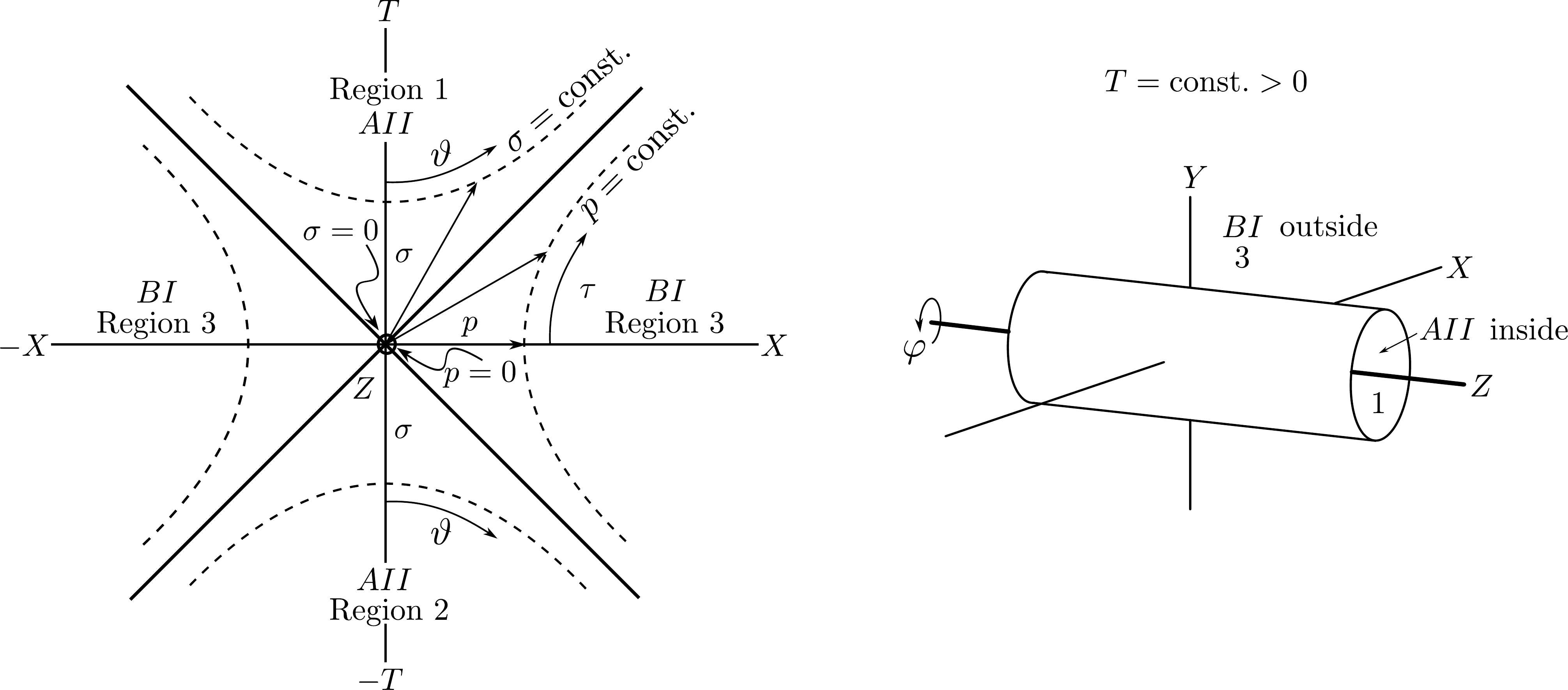}
\caption{Complete Minkowski spacetime divided into Regions~1 and~2 covered by two ``background'' $AII$-metrics
(Region~1 for ${T>0}$ and Region~2 for ${T<0}$ ) and Region 3 covered by the ``background'' $BI$-metric.
The left part is the $T$-$X$ section with ${Y=0}$, $Z$ arbitrary, while the right part shows
the ${X, Y, Z}$ subspace with ${T=\,}$const.${>0}$. For ${T>0}$ the boundary is an expanding cylinder
whose interior is Region~1, while for ${T<0}$ it is a contracting
cylinder whose interior is Region~2. Outside this cylinder lies
Region~3 covered by the $BI$-metric. The tachyon moves along
the $Z$-axis with infinite speed.} \label{img:TachBck}
\end{center}
\end{figure}

As suggested already by  Gott \cite{Gott74}, in the weak-field
limit ${M \to 0}$ the curved spacetime around the tachyonic
source becomes flat Minkowski spacetime, with the tachyon
becoming just a \emph{test particle located along the $Z$-axis}
(identical to $z$-axis). The flat spacetime is \emph{divided
into distinct regions that are separated by the cylindrical
surface} ${T^2=X^2+Y^2}$ with $Z$ arbitrary, as shown in
Fig.~\ref{img:TachBck}.

Region~1 and Region~2 are given by ${T^2>X^2+Y^2}$ with ${T>0}$
and ${T<0}$, respectively. These are covered by the metric
(\ref{AIITach}) with ${M=0}$, namely
\begin{eqnarray}\label{AIITachBck}
\dif s^2=\sigma^2(\dif \p^2+\sinh^2\p\,\dif\q^2)+\dif \z^2-\dif \sigma^2\,,
\end{eqnarray}
whose coordinates are related to background Minkowski
coordinates as ${Z=z}$,
\begin{eqnarray}\label{eq:AIIBckCart}
T \rovno \pm \sigma\,\cosh\p\,,\nonumber\\
X \rovno \sigma\,\sinh\p\,\cos\q\,,\\
Y \rovno \sigma\,\sinh\p\,\sin\q\,,\nonumber
\end{eqnarray}
so that ${T^2-X^2-Y^2=\sigma^2}$. Any ${\sigma=\,}$const. is
thus a hyperboloidal surface. Notice however that the
coordinate singularity ${\sigma=0}$ actually corresponds to
\begin{equation}\label{eq:tachtrajectoryMink}
T=0\,,\qquad X=0=Y\,,
\end{equation}
which is just the $Z$-axis, \emph{the tachyon trajectory}.
\newpage

\begin{figure}[h!]
\begin{center}
\includegraphics[width=130mm]{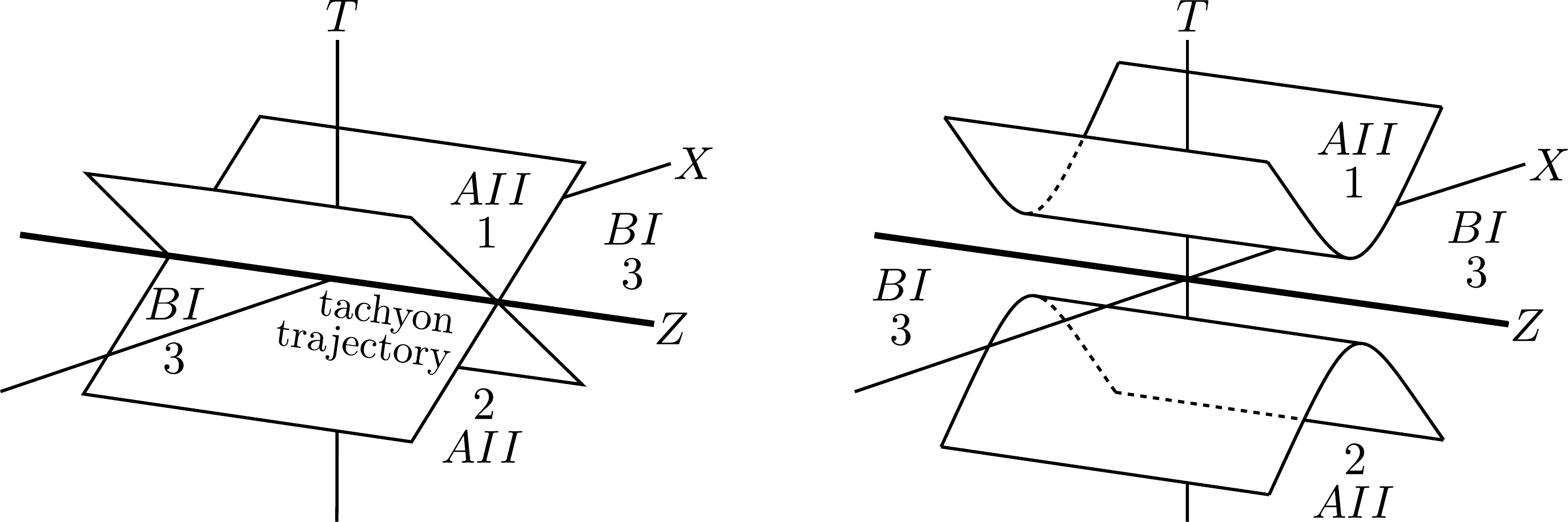}
\caption{The ${T, X, Z}$ subspace of Minkowski spacetime for ${Y=0}$ (left)
and for a constant ${Y\ne0}$ (right). It is divided into two disconnected
Regions~1 and~2 with the $AII$-metrics, and one Region 3 with the $BI$-metric.}
\label{img:TachBckc}
\end{center}
\end{figure}

Region~3 is defined by ${T^2<X^2+Y^2}$. It is covered by the
flat limit of the $BI$-metric (\ref{BITach})
\begin{eqnarray}\label{BITachBck}
\dif s^2=p^2(-\dif \tau^2+\cosh^2\tau\,\dif\q^2)+\dif \z^2+\dif p^2,
\end{eqnarray}
corresponding to Minkowski coordinates via ${Z=z}$,
\begin{eqnarray}\label{eq:BIBckCart}
T \rovno p\,\sinh\tau\,,\nonumber\\
X \rovno p\,\cosh\tau\,\cos\q\,,\\
Y \rovno p\,\cosh\tau\,\sin\q\,,\nonumber
\end{eqnarray}
so that ${X^2+Y^2-T^2=p^2}$. Therefore, ${p=\,}$const. is a
hyperboloidal surface, but the coordinate singularity ${p=0}$
again corresponds to the line ${T=0}$, ${X=0=Y}$, i.e., it is
the tachyon trajectory along the $Z$-axis. It is now also clear
that both regions ${X>0}$ and ${X<0}$ are covered by taking the
full range of the angular coordinate ${\q\in[0,2 \pi)}$.
Moreover, Region~1 for ${Y\not=0}$ is explicitly disconnected
from Region~2, see Fig.~\ref{img:TachBckc}. Thus, two copies of
the metric (\ref{AIITachBck}) together with the single metric
(\ref{BITachBck}) cover the whole Minkowski space (except the
separation boundary ${X^2+Y^2=T^2}$), as shown in
Figs.~\ref{img:TachBck} and~\ref{img:TachBckc}.

\subsection{Curved metrics and their analytic extension} \label{ssc:analext}
Of course, with ${M\ne0}$ the complete spacetime covered by
pairs of (\ref{AIITach}) and (\ref{BITach}) is not flat
anymore. In fact, there is a ``tachyonic-type'' \emph{curvature
singularity} located at ${\sigma=0}$ for the $AII$-metric, and
formally at ${p=0}$ for the $BI$-metric, see
(\ref{BmetricPsi2Lambda}).

Despite the presence of such curvature singularity, both the
metrics \emph{remain asymptotically flat far away from the
tachyonic source}, i.e. for $\sigma$ and $p$ large. Indeed, by
inspecting the Cartesian form of the $AII$-metric
(\ref{eq:AII_cart}) it can be observed that for \emph{any
finite} ${X,Y,Z}$, the metric becomes ${\dif s^2\approx-\dif
T^2+\dif X^2+\dif Y^2+\dif Z^2}$ as ${|T|\to\infty}$. The same
is true for the $BI$-metric (\ref{eq:BI_cart}) for \emph{any
finite} $T$ and ${X^2+Y^2\to\infty}$.

\begin{figure}[t!]
\begin{center}
\includegraphics[width=125mm]{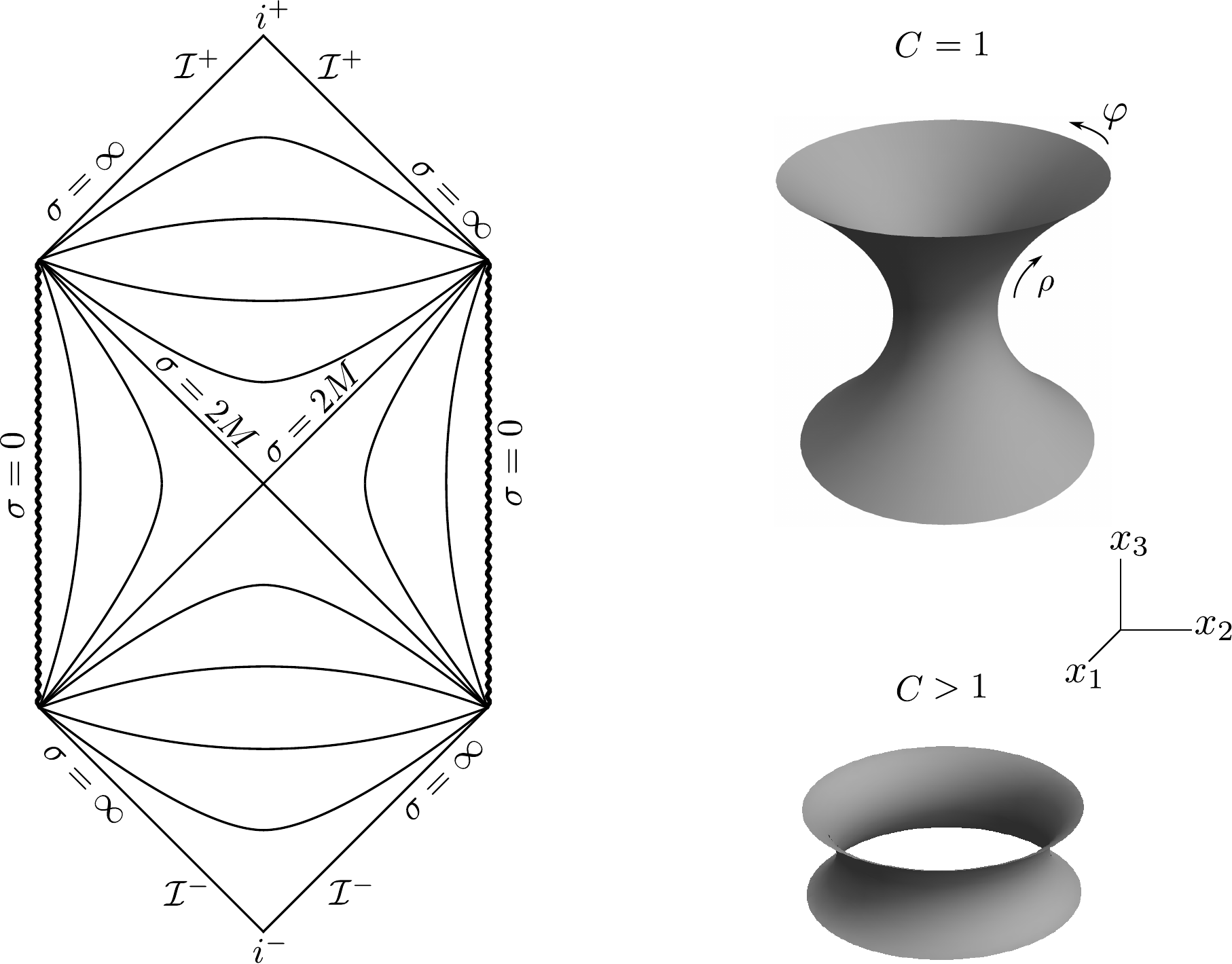}
\caption{Left: Global conformal diagram for the $AII$-metric (\ref{AIITach}) with coordinates ${\p,\q}$
suppressed (in particular, ${\p=0}$). Here $i^{\pm}$ denote past/future timelike infinities,
while $\mathcal{I^{\pm}}$ are null conformal infinities. The horizons ${\sigma=2M}$ are null,
while the singularities ${\sigma=0}$ are timelike.
Right: Embedding diagram of the $BI$-metric (\ref{eq:BIext2}) for section ${z=\rm{const.}}$
and ${\cosh\tau\equiv C=\rm{const.}}$ For larger $C$, the neck diameter grows, while
the surface becomes more restricted. }\label{img:Tach_Penrose}
\end{center}
\end{figure}

There is a coordinate singularity in the $AII$-metric
(\ref{AIITach}) at ${\sigma=2M}$. This is clearly the
\emph{Killing horizon} generated by the Killing vector
$\partial_z$. For ${\sigma>2M}$ the coordinate $z$ is spatial,
and in this region the metric is time-dependent ($\sigma$ is a
temporal coordinate). On the other hand, for ${0<\sigma<2M}$
the coordinate $z$ is temporal, and the spacetime region is
static ($\sigma$ is a spatial coordinate). The $AII$-metric can
be maximally analytically extended across ${\sigma=2M}$,  see
\cite{Gott74} and Section 9.1.1 of \cite{GriPod09} for more
details. The corresponding Penrose conformal diagram can be
constructed by employing the Kruskal--Szekeres-type
coordinates. This is shown in the left part of Fig.~\ref{img:Tach_Penrose}, and
illustrates the null character of the horizons ${\sigma=2M}$,
the timelike character of the curvature singularities
${\sigma=0}$, and asymptotically flat null infinities ${\cal
I}^\pm$ at ${\sigma=\infty}$.

Similarly, the $BI$-metric (\ref{BITach}) with ${M\ne0}$ ceases
to be flat, covering the background exterior region
${T^2<X^2+Y^2}$ via (\ref{eq:BIBckCart}). There is a curvature
singularity at ${p=0}$. However, as explained in
Section~\ref{sbsc:BI}, this singularity \emph{can not be
reached} because only the range ${p\in(2M,\infty)}$ is allowed.
Instead, the $BI$-metric can be analytically extended beyond
${p=2M}$ by performing the transformation ${p=2M /(1-\rho^2)}$
and taking ${\rho\in(-1,1)}$, so that the metric becomes
\begin{equation}\label{eq:BIextensionmetricM}
\dif s^2=4M^2(1-\rho^2\big)^{-4}\big[\big(1-\rho^2\big)^2\big(-\dif\tau^2+\cosh^2\tau\,\dif\q^2\big)
+4\dif\rho^2\big]+\rho^2\dif \z^2\,,
\end{equation}
cf. (\ref{eq:BIextension-1}) and (\ref{eq:BIextensionmetric}).
The two distinct asymptotically
flat regions ${p\to\infty}$ for ${\rho>0}$ and for ${\rho<0}$
are reached as ${\rho\to \pm 1}$. These are joined by a
``bridge/wormhole'' whose ``neck'' is located at ${\rho=0}$,
corresponding to $p=2M$, where the curvature
(\ref{BmetricPsi2Lambda}) is maximal but finite. Its geometry
is ${\dif
s^2_2=4M^2\,(-\dif\tau^2+\cosh^2\tau\,\dif\q^2)}$ which
is a 2-dimensional de~Sitter space.

For fixed values of $\tau$ and $z$, the extended $BI$-metric (\ref{eq:BIextensionmetricM}) reads
\begin{equation}\label{eq:BIext2}
\dif s^2=4M^2(1-\rho^2\big)^{-4}\big[4\dif\rho^2+\big(1-\rho^2\big)^2C^2\dif\q^2
\big]\,,
\end{equation}
where ${C\equiv\cosh\tau \ge 1}$ is a constant. \emph{This
geometry can be embedded} into three-dimensional Euclidean
space with Cartesian coordinates by
\begin{equation}\label{embedding}
x_1 = \frac{2MC}{1-\rho^2}\,\cos\q\,,\qquad
x_2 = \frac{2MC}{1-\rho^2}\,\sin\q\,,\qquad
x_3 = 4M \int\frac{\sqrt{1-C^2\rho^2}}{(1-\rho^2)^2}\,\dif\rho\,.
\end{equation}
For ${\tau=0}$, corresponding to ${C=1}$, we explicitly obtain
${x_3=4M\rho/\sqrt{1-\rho^2}}$. The embedding surface
${x_3^2=8M(\sqrt{x_1^2+x_2^2}-2M)}$, shown in the upper right
part of Fig.~\ref{img:Tach_Penrose}, extends to infinite values
of $x_i$ because the whole range ${\rho\in(-1,1)}$ is allowed.
For ${C>1}$ the integral in (\ref{embedding}) is more
complicated. Numerical integration leads to the axially
symmetric embedding surface shown in the lower right part of
Fig.~\ref{img:Tach_Penrose}. As ${\tau>0}$ (and thus $C$)
grows, the radius $2MC$ of the neck at ${\rho=0}$ grows, while
the allowed range of $\rho$ in (\ref{embedding}) becomes more
restricted to ${|\rho|<1/C}$.

\subsection{Mach--Cherenkov shockwave separating the $AII$ and
$BI$-metrics}\label{sbs:glue}

It has been demonstrated in Section~\ref{ssc:weakfield} that,
in the weak-field limit ${M \to 0}$, the complete Minkowski
spacetime is covered by two ``background'' $AII$-metrics
(Regions~1 and~2 for ${T>0}$ and ${T<0}$, respectively ) and
one ``background'' $BI$-metric (Region 3). These are separated
by the \emph{cylindrical surface} ${X^2+Y^2=T^2}$ with $Z$
arbitrary, see Fig.~\ref{img:TachBck}. This cylinder
\emph{contracts} for ${T<0}$ to the $Z$-axis located at
${T=0}$, which is just the tachyon trajectory, and then
\emph{re-expands} for ${T>0}$. The spacetime region inside this
cylinder is covered by the $AII$-metric, while its exterior is
covered by the $BI$-metric. The cylindrical boundary  between
them, which contracts/expands at the speed of light to/from the
tachyon trajectory, is the \emph{Mach--Cherenkov shockwave
generated by the superluminal source}. Since the tachyon moves
with \emph{infinite speed}, the ``Mach--Cherenkov cone'' is
``infinitely sharp'', i.e., it has a cylindrical geometry.

Of course, with ${M \ne 0}$ the distinct regions covered by the
$AII$ and $BI$ metrics \emph{can not be joined smoothly} across
the Mach--Cherenkov surface. While \emph{keeping the
cylindrical geometry}, it becomes a surface with
\emph{discontinuity} because the specific curvatures on its
both sides are different. This gives rise to a \emph{real
gravitational shockwave}, whose jump in the curvature can be
explicitly evaluated. Instead of using the coordinate
representations (\ref{AIITach}) and (\ref{BITach}), this can be
explicitly performed in the Cartesian coordinates. Notice that
the cylindrical surface ${X^2+Y^2=T^2}$, $Z$ arbitrary,
formally degenerates to ${\sigma=0}$, ${\p=\infty}$ in
(\ref{AIITachBck}) and ${p=0}$, ${\tau=\infty}$ in
(\ref{BITachBck}). By combining the $AII$ and $BI$-metrics in
the Cartesian coordinates (\ref{eq:AII_cart}) and
(\ref{eq:BI_cart}), it is possible to write a
\emph{unified metric for both parts of the curved spacetime}
in the whole range of the background coordinates ${T,X,Y,Z}$
as
\begin{eqnarray}
\label{eq:AII+BIcart1}
\dif s^2 \rovno -\dif T^2+\dif X^2+\dif Y^2+\dif Z^2\\
&&\!\!\!+\frac{2M}{\sqrt{|T^2-X^2-Y^2|}}
\Bigg(-\dif Z^2+\left(1-\frac{2M}{\sqrt{|T^2-X^2-Y^2|}}\right)^{-1}\frac{(-T\dif
T+X\dif X+Y\dif Y)^2}{-T^2+X^2+Y^2}\Bigg).\nonumber
\end{eqnarray}
For ${T^2>X^2+Y^2}$ this is the $AII$-metric (\ref{eq:AII_cart}) with ${M\equiv
n}$, while for ${T^2<X^2+Y^2}$  this is the $BI$-metric (\ref{eq:BI_cart}) with
${M\equiv -n}$.\footnote{It is natural to choose ${M=-n}$ to obtain the
\emph{same} parameter ${M>0}$ for both parts of the unified
metric (\ref{eq:AII+BIcart1}). An alternative choice ${M=n}$
for the $BI$-metric is mathematically also possible.
In such a case the curvature scalar would behave as ${\Psi_2 =
M\left|T^2-X^2-Y^2\right|^{-3/2}}$, but the metric on both
parts would look different.} The metric
(\ref{eq:AII+BIcart1}) diverges on the shock surface
${X^2+Y^2=T^2}$. In fact, there is an \emph{infinite
discontinuity in the Weyl curvature scalar} $\Psi_2$ (\ref{BmetricPsi2Lambda}), namely
\begin{eqnarray}
\Psi_2 \rovno +M\left(T^2-X^2-Y^2\right)^{-3/2}\hspace{11.5mm} \hbox{for}\quad T^2>X^2+Y^2\,,\nonumber\\
\Psi_2 \rovno -M\left(-T^2+X^2+Y^2\right)^{-3/2}\qquad \hbox{for}\quad T^2<X^2+Y^2\,.
\end{eqnarray}
The curvature singularity located at the Mach--Cherenkov
cylindrical shockwave thus has a specific character such that
${\Psi_2\to+\infty}$ when it is approached from its interior,
while ${\Psi_2\to-\infty}$ when it is approached from its
exterior.

\begin{figure}[h!]
\begin{center}
\includegraphics[width=140mm]{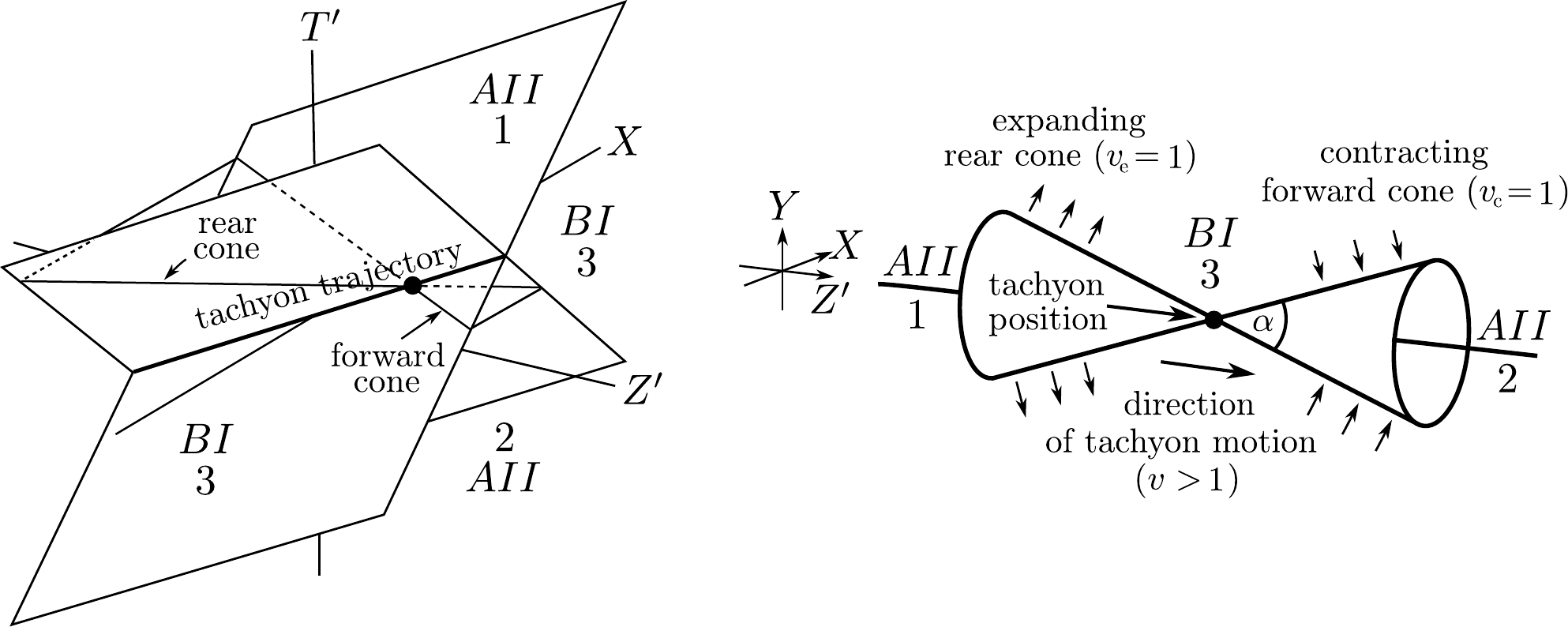}
\caption{Minkowski spacetime separated into Regions~1,~2, and~3 by the Mach-Cherenkov shock cones generated by
the tachyon slowed down by a boost to finite superluminal speed. The spacetime structure is visualized
in sections ${Y=0}$ (left) and ${T'=\,}$const.$>0$ (right). Tachyon moves in the $Z'$-direction,
and generates expanding rear cone and contracting forward cone.}\label{img:TachBckBoost}
\end{center}
\end{figure}

\newpage

\begin{figure}[h!]
\begin{center}
\includegraphics[width=140mm]{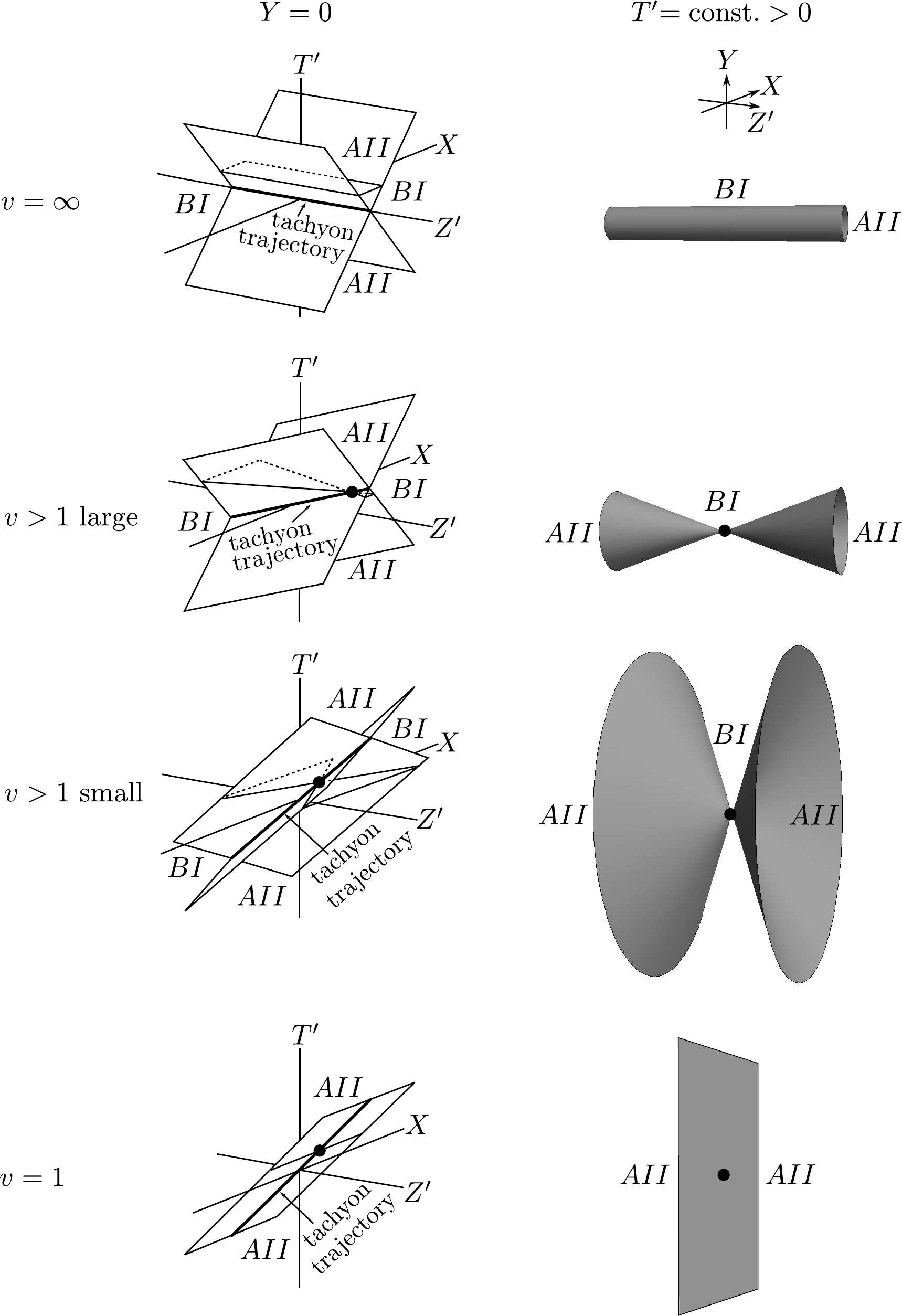}
\caption{The spacetime structure (left), consisting of the $AII$ and $BI$-metrics
separated by  the Mach--Cherenkov shockwaves (right)
for various superluminal speeds $v$ of the tachyonic source.}\label{img:TachBoost}
\end{center}
\end{figure}

\subsection{Boosted metrics}\label{sbsbsc:Boost}
In order to better understand and illustrate the tachyon motion
and also the specific character of the generated
Mach--Cherenkov cone, it is very convenient to consider a boost of
the metric, as originally suggested in \cite{Gott74}.

Since the tachyonic source in (\ref{AIITachBck}) and
(\ref{BITachBck}), and also (\ref{AIITach}) and (\ref{BITach}),
moves at \emph{infinite speed} (it is instantaneously located
everywhere along on the $Z$-axis), such a boost will actually
\emph{slow it down to finite} superluminal speed ${v>1}$. We
will use the boost (\ref{eq:boost_for_AII}). The boosted
tachyon then moves in the $Z'$-direction (coinciding with
${z'}$) at the speed ${v>1}$. In the new coordinates, the
surface ${\sqrt{X^2+Y^2}=T}$ that separates Regions~1 and~3,
and the surface ${\sqrt{X^2+Y^2}=-T}$ that separates Regions~2
and~3, take the form
\begin{equation}\label{eq:Cercone}
X^2+Y^2=\frac{(v\,T'-Z')^2}{v^2-1}\,.
\end{equation}

At any fixed time $T'$, these represent \emph{Mach--Cherenkov
shock cones with the vertex at} ${Z'=v\,T'}$ which is the
\emph{actual position of the tachyon}, and with the \emph{angle}
$\alpha$ of the cone such that
${\alpha=2\,\hbox{arccot}\sqrt{v^2-1}}$. As illustrated in
Fig.~\ref{img:TachBckBoost}, the \emph{rear cone expands} while
the \emph{forward cone contracts} at the speed of light. The
tachyon is always located at the intersection of these cones
and --- in a manner similar to the so-called scissors effect
--- moves faster than light.

It is also illustrative to plot these shockwaves for
\emph{different speeds} ${v>1}$ of the tachyon, see
Fig.~\ref{img:TachBoost}. For larger superluminal $v$, the
angle $\alpha$ of the cone is smaller. In the extreme case
${v=\infty}$ this angle is zero and the cone degenerates to the
cylinder plotted in Fig.~\ref{img:TachBck}, while in the
opposite limit ${v=1}$ both the rear and front cones coalesce
and form a shock plane propagating at the speed of light along
$Z'$. This behavior follows from the dependence of the regions
on $v$. The interior Regions~1 and~2 covered by two separate
$AII$-metrics are located at ${X^2+Y^2<(v\,T'-Z')^2/(v^2-1)}$,
so that \emph{for smaller superluminal speed} $v$ these regions
are \emph{larger}. In the limit ${v\to1}$ the whole spacetime
(except the Mach--Cherenkov shock, now located at ${Z'=T'}$) is
covered by the pair of $AII$-metrics. On the other hand,
Region~3 defined by ${X^2+Y^2>(v\,T'-Z')^2/(v^2-1)}$ becomes
\emph{smaller}, and in the limit ${v\to1}$ it disappears.

\section{$AII$ and $BI$-metrics with $\Lambda$ are the Schwarzschild--(anti-)de~Sitter spacetime boosted to infinite speed}
\label{sc:boostLambda}

In previous sections we considered the $A$ and $B$-metrics in
Minkowski background by setting ${\Lambda=0}$ in the metrics
(\ref{AmetricLambda}) and (\ref{BmetricLambda}), respectively.
Now we are going to extend the results to any value of the
cosmological constant. In fact, we will demonstrate that these
metrics can be understood as specific regions of the spacetime
representing exact gravitational field of a tachyonic source
moving in de~Sitter (if ${\Lambda>0}$) or anti--de~Sitter (if
${\Lambda<0}$) universe.

\begin{figure}[t!]
\begin{center}
\includegraphics[width=110mm]{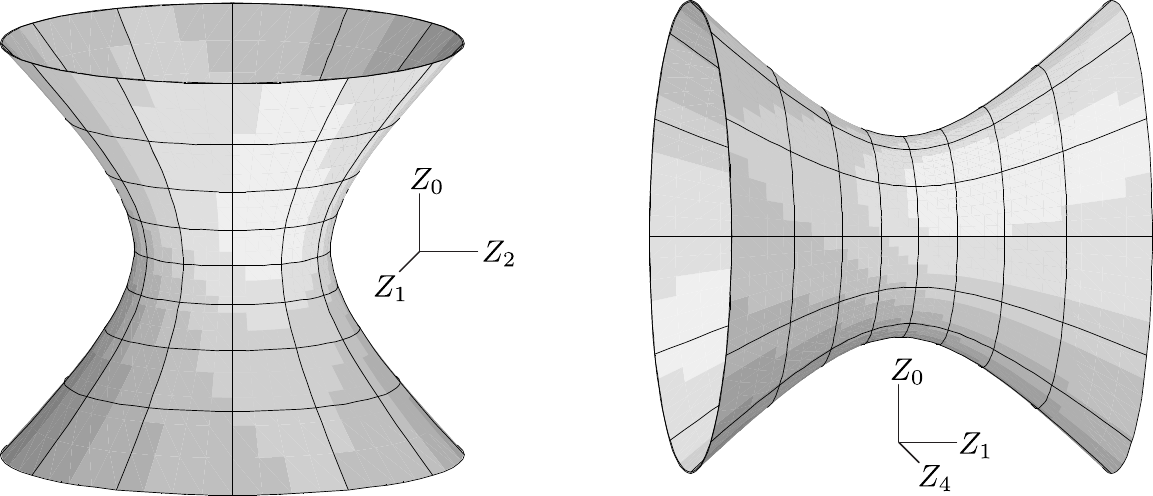}
\caption{The visualizations of de~Sitter spacetime (left) and anti-de~Sitter spacetime (right)
as hyperboloids (\ref{eq:hyperboloids}) embedded in a flat five-dimensional spacetime (\ref{eq:5D}).
The remaining coordinates ($Z_3, Z_4$ for ${\Lambda>0}$, and $Z_2,Z_3$ for ${\Lambda<0}$)
are suppressed. For more details see~\cite{GriPod09}.
}\label{img:Hyperboloids}
\end{center}
\end{figure}

First, let us investigate boosts of the classic
Schwarzschild--(anti-)de~Sitter metric (which is the most
important $AI$-metric with $\Lambda$) and perform the limit
${v\rightarrow \infty}$. However, with ${\Lambda\neq0}$ the
background is not flat but it is everywhere curved
(anti-)de~Sitter spacetime. To perform the boost correctly, it
is most convenient to employ a five-dimensional embedding
formalism. It is well known that (anti-)de~Sitter spacetime can
be understood as a hyperboloid
\begin{equation}\label{eq:hyperboloids}
-Z_0^2+Z_1^2+Z_2^2+Z_3^2+\e Z_4^2=\e\,a^2\,,
\end{equation}
embedded into a five-dimensional flat spacetime
\begin{equation}\label{eq:5D}
\dif s^2=-\dif Z_0^2+\dif Z_1^2+\dif Z_2^2+\dif Z_3^2+\e\,\dif Z_4^2\,,
\end{equation}
where ${a\equiv\sqrt{3/|\Lambda|}}$ and ${\e\equiv\sign
\Lambda}$, see the visualizations in
Fig.~\ref{img:Hyperboloids}.

The (anti-)de~Sitter background is obtained from the
$AI$-metric (\ref{AmetricLambda}) for
${\epsilon_2=1=\epsilon_0}$, ${n=0}$. In the case
${\Lambda>0}$, these coordinates parametrize the hyperboloid
(\ref{eq:hyperboloids}) as
\begin{equation}\label{eq:AI_dS_par}
\left. \begin{array}{l}
Z_0 = \pm\sqrt{a^2-p^2}\,\sinh(t/a)\,, \\
Z_1 = p\,\sqrt{1-q^2}\,\cos\q\,, \\
Z_2 = p\,\sqrt{1-q^2}\,\sin\q\,, \\
Z_3 = p\,q\,, \\
Z_4 = \pm\sqrt{a^2-p^2}\,\cosh(t/a)\,,
 \end{array} \!\right\} \ \textrm{for}\ p<a, \ \left. \
 \begin{array}{l}
Z_0 = \pm\sqrt{p^2-a^2}\,\cosh(t/a)\,, \\
Z_1 = p\,\sqrt{1-q^2}\,\cos\q\,, \\
Z_2 = p\,\sqrt{1-q^2}\,\sin\q\,, \\
Z_3 = p\,q\,, \\
Z_4 = \pm\sqrt{p^2-a^2}\,\sinh(t/a)\,,
 \end{array} \right\}\ \textrm{for}\ p>a\,,
\end{equation}
while for ${\Lambda<0}$ the corresponding parametrization is
\begin{eqnarray}\label{eq:AI_adS_par}
Z_0\rovno\sqrt{a^2+p^2}\,\sin (t/a)\,,\nonumber\\
Z_1\rovno p\,\sqrt{1-q^2}\,\cos\q\,,\nonumber\\
Z_2\rovno p\,\sqrt{1-q^2}\,\sin\q\,,\\
Z_3\rovno p\,q\,,\nonumber\\
Z_4\rovno \sqrt{a^2+p^2}\,\cos(t/a)\,.\nonumber
\end{eqnarray}
Expressing the $AI$-metric (\ref{AmetricLambda}) with
${\epsilon_2=1=\epsilon_0}$ and ${n\neq 0}$ in these
five-dimensional coordinates, using (\ref{eq:AI_dS_par}),
(\ref{eq:AI_adS_par}), we obtain
\begin{equation}\label{eq:AILambda5D}
\dif s^2=\dif s^2_{\rm (A)dS}+\frac{2m a^2}{p}\left(\frac{(Z_4\dif Z_0-Z_0\dif Z_4)^2}{(Z_4^2-\e Z_0^2)^2}
+\frac{a^2}{p^2}\frac{(Z_0\dif Z_0-\e Z_4\dif Z_4)^2}{(Z_4^2-\e Z_0^2)(Z_4^2-\e Z_0^2-2ma^2/p)}\right),
\end{equation}
where $\dif s^2_{\rm (A)dS}$ is the (anti-)de~Sitter background
metric (\ref{eq:5D}), ${m\equiv-n}$ and
${p=\sqrt{Z_1^2+Z_2^2+Z_3^2}}$. As in
\cite{HottaTanaka93,PodGri97,PodGri98}, we can make a boost
similar to (\ref{eq:boost_for_Schw}), but now in the
coordinates $Z_0, Z_3$:
\begin{equation}\label{eq:boost_for_SchwdS}
Z_0=\frac{Z_0'+v\,Z_3'}{\sqrt{1-v^2}}\,,\qquad Z_3=\frac{Z_3'+v\,Z_0'}{\sqrt{1-v^2}}\,.
\end{equation}
We immediately observe that all the terms introducing the
factor $\sqrt{1-v^2}$ into (\ref{eq:AILambda5D}) are
\emph{quadratic}, so it is possible to make the formal limit
${v\to\infty}$ that will effectively cause just a swap
${Z_0^2\rightarrow -Z_3'^2}$ and ${Z_3^2\rightarrow -Z_0'^2}$
in (\ref{eq:AILambda5D}). The resulting metric will thus become
\begin{equation}\label{eq:AILambda_inf_boost}
\dif s^2=\dif s^2_{\rm (A)dS}+\frac{2m a^2}{p}\left(\frac{-(Z_4\dif Z_3'-Z_3'\dif Z_4)^2}{(Z_4^2+\e Z_3'^2)^2}
+\frac{a^2}{p^2}\frac{(Z_3'\dif Z_3'+\e Z_4\dif Z_4)^2}{(Z_4^2+\e Z_3'^2)(Z_4^2+\e Z_3'^2-2ma^2/p)}\right),
\end{equation}
where ${p=\sqrt{Z_1^2+Z_2^2-Z_0'^2}}$. This is the $AII$-metric
in the region ${Z_0'^2>Z_1^2+Z_2^2}$ with a purely imaginary
mass, and the $BI$-metric in the region ${Z_0'^2<Z_1^2+Z_2^2}$
with a real mass.

Indeed, the $AII$-metric background (\ref{AmetricLambda}) for
${\epsilon_2=-1=\epsilon_0}$, ${n=0}$, with ${\Lambda>0}$ is
given by
\begin{eqnarray}\label{eq:AII_dS_par}
Z_0\rovno p\,q\,,\nonumber\\
Z_1\rovno p\,\sqrt{q^2-1}\,\cos\q\,,\nonumber\\
Z_2\rovno p\,\sqrt{q^2-1}\,\sin\q\,,\\
Z_3\rovno \sqrt{p^2+a^2}\,\cos(t/a)\,,\nonumber\\
Z_4\rovno \sqrt{p^2+a^2}\,\sin(t/a)\,,\nonumber
\end{eqnarray}
and with ${\Lambda<0}$
\begin{equation}\label{eq:AII_adS_par}
\left. \begin{array}{l}
Z_0 = p\,q\,, \\
Z_1 = p\,\sqrt{q^2-1}\,\cos\q\,, \\
Z_2 = p\,\sqrt{q^2-1}\,\sin\q\,, \\
Z_3 = \pm\sqrt{a^2-p^2}\,\sinh(t/a)\,, \\
Z_4 = \pm\sqrt{a^2-p^2}\,\cosh(t/a)\,,
 \end{array} \!\right\} \ \textrm{for}\ {p<a\,,} \ \left. \
 \begin{array}{l}
Z_0 = p\,q\,, \\
Z_1 = p\,\sqrt{q^2-1}\,\cos\q\,, \\
Z_2 = p\,\sqrt{q^2-1}\,\sin\q\,, \\
Z_3 = \pm\sqrt{p^2-a^2}\,\cosh(t/a)\,, \\
Z_4 = \pm\sqrt{p^2-a^2}\,\sinh(t/a)\,,
 \end{array} \right\}\ \textrm{for}\ {p>a\,.}
\end{equation}
These parametrizations  only cover the region
${Z_0^2>Z_1^2+Z_2^2}$. The complete $AII$-metric with ${n \ne
0}$ written in the coordinates of (\ref{eq:AII_dS_par}),
(\ref{eq:AII_adS_par}) thus has the form
\begin{eqnarray}
\dif s^2=\dif s^2_{\rm (A)dS}+\frac{2n a^2}{p}\left(\frac{-(Z_4\dif Z_3-Z_3\dif Z_4)^2}{(Z_4^2+\e Z_3^2)^2}
-\frac{a^2}{p^2}\frac{(Z_3\dif Z_3+\e Z_4\dif Z_4)^2}{(Z_4^2+\e Z_3^2)(Z_4^2+\e Z_3^2-2na^2/p)}\right),
\end{eqnarray}
where ${p=\sqrt{Z_0^2-Z_1^2-Z_2^2}}$. We clearly see that this
is exactly the Schwarzschild--(anti-)de~Sitter metric
(\ref{eq:AILambda_inf_boost}) boosted to infinite speed, with
the identification ${m=\im\,n}$ (and ${p^2 \to -p^2}$).

Similarly, the $BI$-metric (\ref{BmetricLambda}) for
${\epsilon_2=1}$, ${\epsilon_0=-1}$, ${n=0}$ corresponds to the
parametrization
\begin{equation}\label{eq:BI_adS_par}
\left. \begin{array}{l}
Z_0 = p\,q\,, \\
Z_1 = p\,\sqrt{1+q^2}\,\cos t\,, \\
Z_2 = p\,\sqrt{1+q^2}\,\sin t\,, \\
Z_3 = \sqrt{a^2-p^2}\,\cos(z/a)\,, \\
Z_4 = \sqrt{a^2-p^2}\,\sin(z/a)\,,
 \end{array} \!\right\} \ \textrm{for}\ {\Lambda>0\,,} \ \left. \
 \begin{array}{l}
Z_0 = p\,q\,, \\
Z_1 = p\,\sqrt{1+q^2}\,\cos t\,, \\
Z_2 = p\,\sqrt{1+q^2}\,\sin t\,, \\
Z_3 = \pm\sqrt{a^2+p^2}\,\sinh(z/a)\,, \\
Z_4 = \pm\sqrt{a^2+p^2}\,\cosh(z/a)\,,
 \end{array} \right\}\ \textrm{for}\ {\Lambda<0\,,}
\end{equation}
which only covers the region ${Z_0^2<Z_1^2+Z_2^2}$. In terms of
the coordinates (\ref{eq:BI_adS_par}), the complete $BI$-metric
reads
\begin{equation}
\dif s^2=\dif s^2_{\rm (A)dS}+\frac{2n a^2}{p}\left(\frac{(Z_4\dif Z_3'-Z_3'\dif Z_4)^2}{(Z_4^2+\e Z_3'^2)^2}
-\frac{a^2}{p^2}\frac{(Z_3'\dif Z_3'+\e Z_4\dif Z_4)^2}{(Z_4^2+\e Z_3'^2)(Z_4^2+\e Z_3'^2+2na^2/p)}\right),
\end{equation}
which is again the same as the Schwarzschild--de~Sitter metric
boosted to infinite speed (\ref{eq:AILambda_inf_boost}) with
${m=-n}$ and ${p=\sqrt{Z_1^2+Z_2^2-Z_0^2}}$.

We may thus conclude that \emph{both} the $AII$ and
$BI$-metrics with any $\Lambda$ can be understood as (formal)
\emph{limits of the classic Schwarzschild--(anti-)de~Sitter
metric boosted to infinite speed}. Of course, complementary
procedures can also be applied: the Schwarzschild-de~Sitter
metric can be obtained by slowing down (stopping) the source of
the $AII$-metric with imaginary mass, or the $BI$-metric with
real mass.

\section{Coordinate ranges and extensions of the $B$-metrics with ${\Lambda\ne0}$}\label{sc:rangesLambda}
To understand the global character of the $B$-metrics with any
cosmological constant, and their possible extensions and
combinations, it is necessary to analyze the admitted
coordinate ranges.

\subsection{$BI$-metric with $\Lambda$}\label{sbsc:BILambda}
\begin{figure}[ht!]
\begin{center}
\includegraphics[width=150mm]{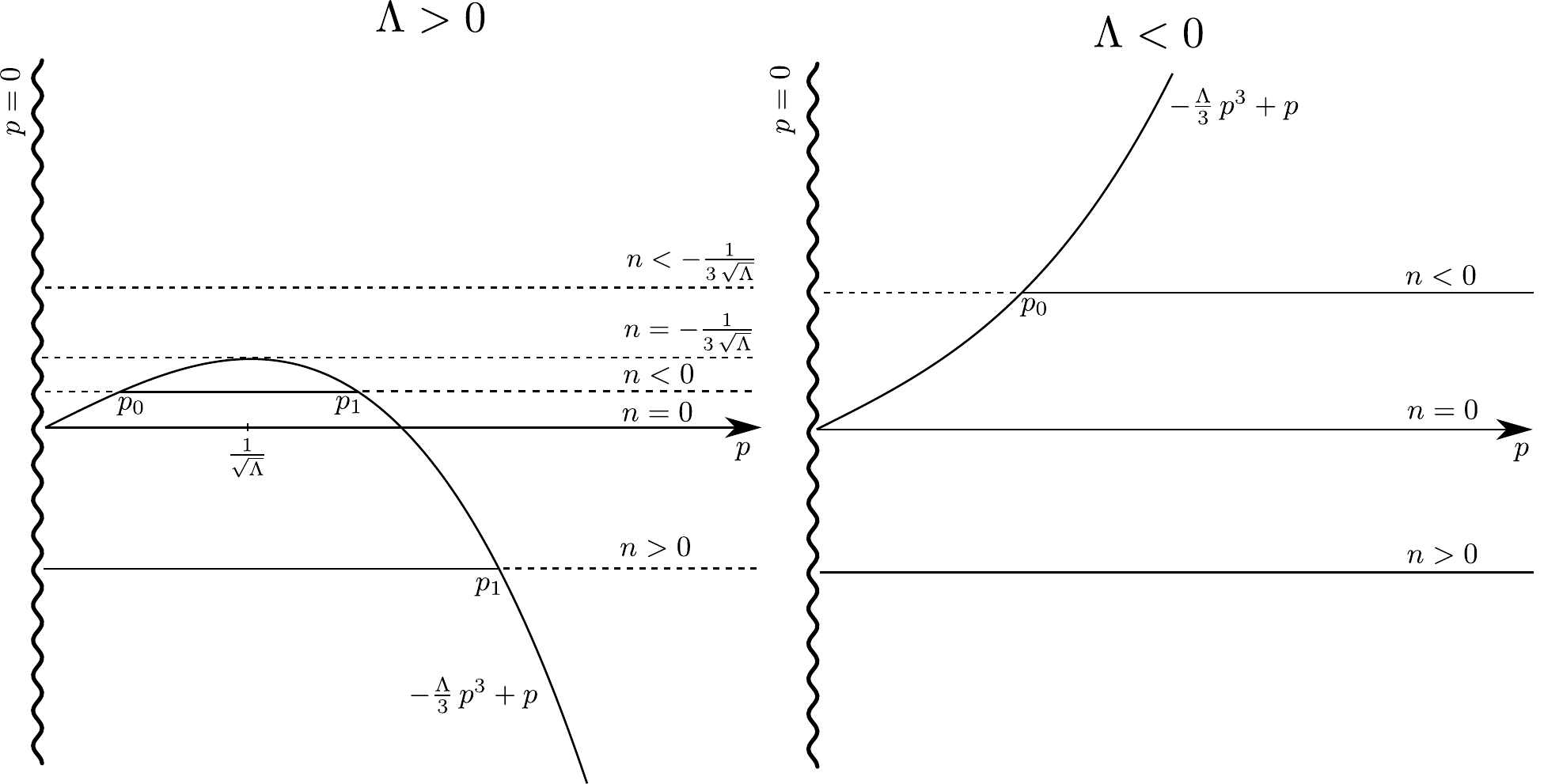}
\caption{
Allowed ranges of the coordinate $p$ for ${\Lambda>0}$ (left) and ${\Lambda<0}$ (right)
in the $BI$-metric (\ref{BImetricLambda}). The horizontal lines correspond to different
values of ${-2\n}$. The parts of these lines that lie under the curve ${-\frac{\Lambda}{3}\,p^3+p}$
determine the admitted range of $p$. Their parts above
the curve are dashed --- there is no solution for these values
because the metric would no longer have the correct signature.  The intersections
mark the roots $p_0$, $p_1$ of (\ref{eq:BIcubic}), with ${0<p_0<p_1}$.
}\label{img:BIRangesofp}
\end{center}
\end{figure}

For ${\epsilon_2=1}$ the metric (\ref{BmetricLambda}) is
\begin{equation}
\dif s^2 = -p^2(\epsilon_0-q^2)\,\dif t^2
 +\frac{p^2}{\epsilon_0-q^2}\,\dif q^2
 +\Big(1+\frac{2\n}{p}-\frac{\Lambda}{3}\,p^2\Big)\dif \z^2
 +\Big(1+\frac{2\n}{p}-\frac{\Lambda}{3}\,p^2\Big)^{-1}\dif p^2\,.
\label{BImetricLambda}
\end{equation}
The coordinate ranges are (considering ${\epsilon_0=-1}$)
${q,\z\in\mathbb{R}}$, ${t\in[0,2\pi)}$, while the range of $p$
depends on $\n$ and $\Lambda$. It is determined by the roots of
the cubic equation
\begin{equation}\label{eq:BIcubic}
-\frac{\Lambda}{3}\,p^3+p+2\n=0\,.
\end{equation}
The best way to illustrate the allowed ranges of ${p>0}$ for
all possible cases is to plot the possible roots $p_i$ of
(\ref{eq:BIcubic}) as intersections of the function
${-\frac{\Lambda}{3}\,p^3+p}$ with horizontal lines
corresponding to various values of ${-2\n}$. Since the metric
coefficients $g_{zz}$ and $g_{pp}$ must remain positive, we
require ${-\frac{\Lambda}{3}\,p^3+p>-2\n}$. Explicit
visualization is given in Fig.~\ref{img:BIRangesofp}. The left
part applies to ${\Lambda>0}$, the right part to ${\Lambda<0}$.
It can be seen that for ${\Lambda>0}$ and ${\n\geq0}$, the
values of $p$ are ${p \in (0, p_1)}$. For
${-\frac{1}{3\,\sqrt{\Lambda}}<\n<0}$, the allowed range is ${p
\in (p_0,p_1)}$ given by two roots of (\ref{eq:BIcubic}). For
${\n=-\frac{1}{3\,\sqrt{\Lambda}}}$ the metric degenerates
because only one value ${p=\frac{1}{\sqrt{\Lambda}}}$ is
allowed. Finally, there is no solution for
${\n<-\frac{1}{3\,\sqrt{\Lambda}}}$. In the case ${\Lambda<0}$,
the situation is much simpler because the function
${-\frac{\Lambda}{3}\,p^3+p}$ monotonously grows from zero.
Therefore, for ${\n\geq0}$ the coordinate $p$ takes the maximal
range $(0,\infty)$, while for ${\n<0}$ its range is restricted
to ${p \in (p_0,\infty)}$.

Interestingly, it is possible to analytically extend the
$BI$-metric (\ref{BImetricLambda}) across $p_i$ (the roots of
$g_{zz}$) by performing the transformation
\begin{eqnarray}\label{eq:BILambdaextension}
\rho^2=1+\frac{2\n}{p}-\frac{\Lambda}{3}p^2\,,\qquad \sinh\tau=q\,,\qquad
\zeta=\frac{1}{2\n}\,z\,,\qquad \q=t\,.
\end{eqnarray}
In fact, for ${\Lambda = 0}$ this reduces to
(\ref{eq:BIextension-1}). The resulting metric is
\begin{equation}\label{eq:BILambdaextensionmetric}
\dif s^2=4\n^2\Big[R_1(\rho)\big(-\dif\tau^2+\cosh^2\tau\,\dif\q^2\big)+\rho^2\dif \zeta^2
+R_2(\rho)\,\dif\rho^2\Big]\,,
\end{equation}
where ${R_1(\rho)=\frac{1}{4n^2}\,p^2(\rho)}$,
${R_2(\rho)=\frac{1}{4n^2}\,\big(\n\,p(\rho)^{-2}+\frac{1}{3}\Lambda\,p(\rho)\big)^{-2}}$,
with $p(\rho)$ obtained by inverting the relation
(\ref{eq:BILambdaextension}). For ${\Lambda = 0 }$ we recover
the metric (\ref{eq:BIextensionmetric}). Relations between the
ranges of $p$ and the ranges of ${\rho>0}$ are shown in
Table~\ref{tbl:BIprange}. Analytic extension across ${\rho=0}$
is obtained by admitting a ``mirror chart'' with ${\rho<0}$.

\begin{table}[h]
\begin{center}
\begin{tabular}{| c | c || c | c |}
\hline
$\Lambda$ & $n$ & range of $p$ & range of $\rho>0$ \\
\hline
\hline
$>0$ & $>0$ & $(0,p_1)$ & $(0,\infty)$\\
\hline
$>0$ & $<0$ & $(p_0,p_1)$ & $(0,\rho_{\textrm{max}})$\\
\hline
$<0$ & $>0$ & $(0,\infty)$ & $(\rho_{\textrm{min}},\infty)$\\
\hline
$<0$ & $<0$ & $(p_0,\infty)$ & $(0,\infty)$\\
\hline
\end{tabular}
\caption{Ranges of $p$ and the corresponding ranges of ${\rho}$
for possible combinations of $\Lambda$ and~$\n$ in the
$BI$-metric~(\ref{BImetricLambda}). Here $\rho_{\textrm{min}}$
and $\rho_{\textrm{max}}$ denote specific minimal and maximal
values or $\rho$.}\label{tbl:BIprange}
\end{center}
\end{table}

\newpage

\subsection{$BII$-metric with $\Lambda$}\label{sbsc:BIILambda}
\begin{figure}[h!]
\begin{center}
\includegraphics[width=150mm]{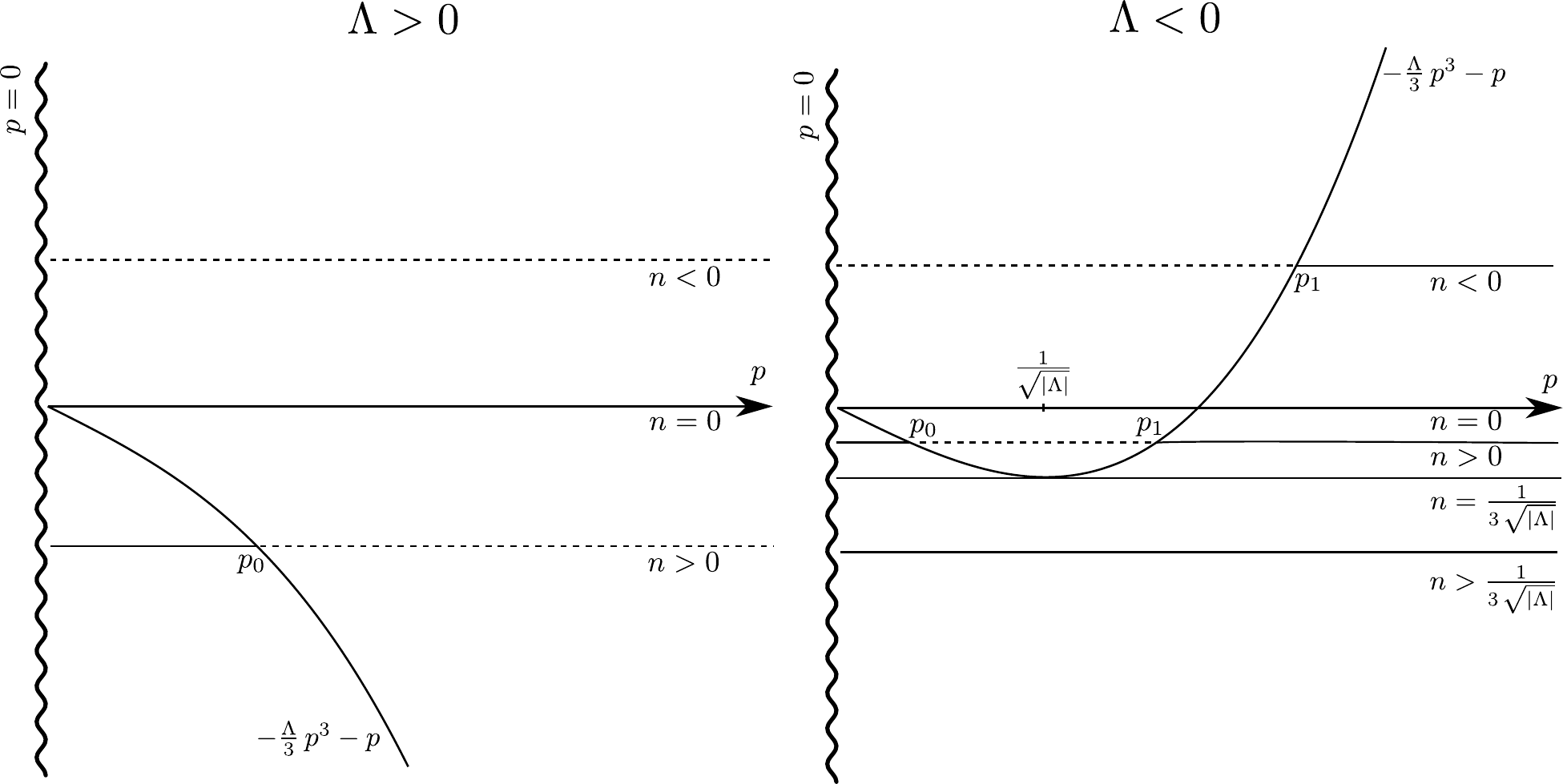}
\caption{
Allowed ranges of $p$ for ${\Lambda>0}$ (left) and ${\Lambda<0}$ (right)
in the $BII$-metric (\ref{BIImetricLambda}). They are determined by those parts
of the horizonal lines ${-2\n}$ that lie under the curve ${-\frac{\Lambda}{3}\,p^3-p}$.
}\label{img:BIIRangesofp}
\end{center}
\end{figure}

For ${\epsilon_2=-1}$ the metric (\ref{BmetricLambda}) gives
\begin{equation}
\dif s^2= -p^2(\epsilon_0+q^2)\,\dif t^2
 + \frac{p^2}{\epsilon_0+q^2}\,\dif q^2
 +\Big(-1+\frac{2n}{p}-\frac{\Lambda}{3}\,p^2\Big)\dif \z^2
 +\Big(-1+\frac{2n}{p}-\frac{\Lambda}{3}\,p^2\Big)^{-1}\dif p^2\,,
\label{BIImetricLambda}
\end{equation}
where ${t,q,\z\in\mathbb{R}}$  (considering ${\epsilon_0=1}$).
The range of $p$ is given by the roots ${0<p_0<p_1}$ of
\begin{equation}\label{eq:BIIcubic}
-\frac{\Lambda}{3}\,p^3-p+2n=0\,.
\end{equation}
We plot the allowed ranges of $p$ for different values of $\Lambda$
and $n$ in Fig.~\ref{img:BIIRangesofp}. For
${\Lambda>0}$, ${n>0}$ the range is ${p \in (0, p_0)}$, while
the case ${n\leq0}$ is forbidden. For ${\Lambda<0}$ the
coordinate $p$ has the full range $(0,\infty)$ for any
${n>\frac{1}{3\,\sqrt{{|\Lambda|}}}}$. If
${\n=\frac{1}{3\,\sqrt{|\Lambda|}}}$ the metric degenerates at
${p=\frac{1}{\sqrt{|\Lambda|}}}$. If
${0<n<\frac{1}{3\,\sqrt{|\Lambda|}}}$, the metric represents
two separate regions ${p\in(0,p_0)}$ and ${p\in(p_1,\infty)}$.
Finally, if ${n\leq 0}$ the coordinate $p$ can only take values
${p \in (p_1,\infty)}$. Notice that ${n\leq0}$ is forbidden for
the $BII$-metric (\ref{eq:BIImetric}) with ${\Lambda=0}$ and
(\ref{BIImetricLambda}) with ${\Lambda>0}$. In particular, the
$BII$-metric (\ref{BIImetricLambda}) with ${\Lambda<0}$ (and
sufficiently large ${|\Lambda|}$) includes the anti--de~Sitter
spacetime when ${n=0}$.

Again, we can perform its analytic extension across $p_i$, in
this case by generalizing (\ref{eq:BIIextension}) to
\begin{eqnarray}\label{eq:BIILambdaextension}
\rho^2=-1+\frac{2\n}{p}-\frac{\Lambda}{3}p^2\,,\qquad \sinh z=q\,,
\qquad \zeta=\frac{1}{2\n}\,z\,.
\end{eqnarray}
The extended $BII$-metric will then be
\begin{equation}\label{eq:BIILambdaextensionmetric}
\dif s^2=4\n^2\Big[R_1(\rho)\big(\dif z^2-\cosh^2z\,\dif t^2\big)+\rho^2\dif \zeta^2
+R_2(\rho)\,\dif\rho^2\Big]\,.
\end{equation}
Explicit form of the metric functions ${R_1(\rho), R_2(\rho)}$
(reducing to (\ref{eq:BIIextensionmetric}) when ${\Lambda=0}$)
is obtained by using the function $p(\rho)$ that is obtained by
inverting expression (\ref{eq:BIILambdaextension}). The ranges
of ${\rho>0}$ corresponding to the allowed ranges of $p$ are
shown in Table~\ref{tbl:BIIprange}.

\begin{table}[h]
\begin{center}
\begin{tabular}{| c | c || c | c |}
\hline
$\Lambda$ & $n$ & range of $p$ & range of $\rho>0$\\
\hline
\hline
$>0$ & $>0$ & $(0,p_0)$ & $(0,\infty)$\\
\hline
$<0$ & $>\frac{1}{3\,\sqrt{|\Lambda|}}$ & $(0,\infty)$ & $(\rho_{\textrm{min}},\infty)$\\
\hline
$<0$ & $0<n<\frac{1}{3\,\sqrt{|\Lambda|}}$ & $(0,p_0),(p_1,\infty)$ &
$(0,\infty)$\\
\hline
$<0$ & $<0$ & $(p_1,\infty)$ & $(0,\infty)$\\
\hline
\end{tabular}
\caption{Ranges of $p$ and $\rho$ for possible  $\Lambda$ and
$\n$ in the
$BII$-metric~(\ref{BIImetricLambda}).}\label{tbl:BIIprange}
\end{center}
\end{table}

\newpage
\subsection{$BIII$-metric with $\Lambda$}\label{sbsc:BIIILambda}
\begin{figure}[h!]
\begin{center}
\includegraphics[width=150mm]{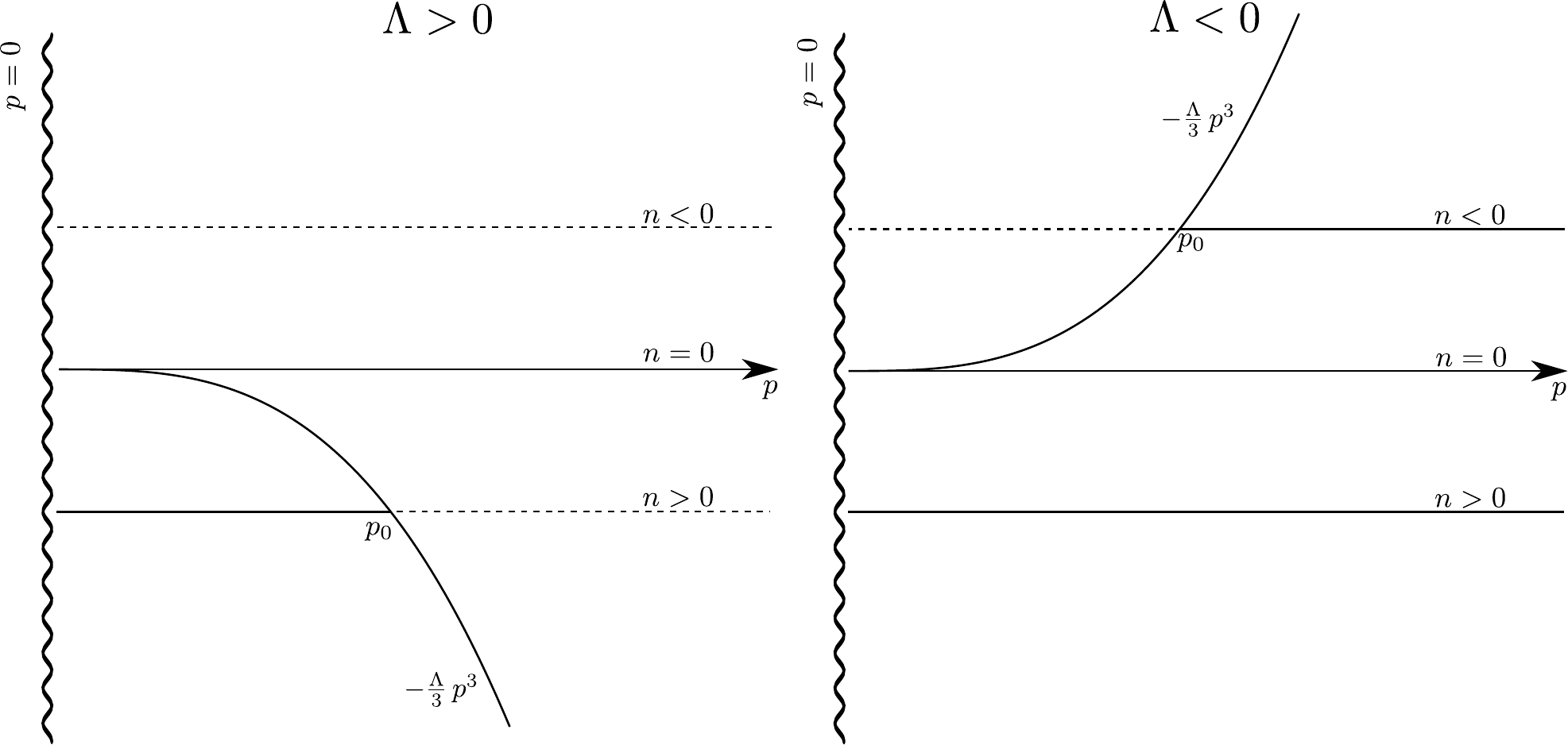}
\caption{
Allowed ranges of $p$ for ${\Lambda>0}$ (left) and ${\Lambda<0}$ (right)
in the $BIII$-metric (\ref{BIIImetricLambda}). They are determined by
the root ${p_0=\sqrt[3]{6n/\Lambda}}$ of (\ref{eq:BIIIcubic}).
}\label{img:BIIIRangesofp}
\end{center}
\end{figure}

The metric (\ref{BmetricLambda}) for ${\epsilon_2=0}$ (and,
without loss of generality, ${\epsilon_0=1}$) reads
\begin{equation}\label{BIIImetricLambda}
\dif s^2= -p^2\,\dif t^2
 + p^2\,\dif q^2
 +\Big(\frac{2n}{p}-\frac{\Lambda}{3}\,p^2\Big)\dif \z^2
 +\Big(\frac{2n}{p}-\frac{\Lambda}{3}\,p^2\Big)^{-1}\dif p^2\,,
\end{equation}
where $t,q,z\in\mathbb{R}$, and the range of $p$ is determined
by the roots of
\begin{eqnarray}\label{eq:BIIIcubic}
-\frac{\Lambda}{3}p^3+2n=0\,.
\end{eqnarray}
The results are visualized in Fig.~\ref{img:BIIIRangesofp}. For
${\Lambda>0}$, ${n>0}$, the allowed range is ${p\in(0,p_0)}$.
As for the $BII$-metric, the case ${n\leq0}$ is not allowed.
When ${\Lambda<0}$, ${n\geq0}$, the coordinate $p$ take the
whole range ${(0,\infty)}$, while for ${n< 0}$, it takes
$(p_0,\infty)$. Of course, equation (\ref{eq:BIIIcubic}) can be
explicitly solved: For ${n/\Lambda>0}$, the root is
${p_0=\sqrt[3]{6n/\Lambda}}$, while for ${n/\Lambda<0}$ there
is no positive root. These $BIII$-metrics are, in fact,
equivalent to the Linet--Tian metric, see \cite{GPLT,
PodHruGri18}.

\begin{table}[h]
\begin{center}
\begin{tabular}{| c | c || c | c |}
\hline
$\Lambda$ & $n$ & range of $p$ & range of $\rho>0$\\
\hline
\hline
$>0$ & $>0$ & $(0,p_0)$ & $(0,\infty)$\\
\hline
$<0$ & $>0$ & $(0,\infty)$ & $(\rho_{\textrm{min}},\infty)$\\
\hline
$<0$ & $<0$ & $(p_0,\infty)$ & $(0,\infty)$\\
\hline
\end{tabular}
\caption{Ranges of $p$ and $\rho$ for possible  $\Lambda$ and
$\n$ in the $BIII$-metric~(\ref{BIIImetricLambda}).}\label{tbl:BIIIprange}
\end{center}
\end{table}

We can also make an analytic extension beyond $p_0$ (which does not have a counterpart for the ${\Lambda=0}$ case because there are no roots of $g_{zz}$). Such an extension is achieved by
\begin{eqnarray}\label{eq:BIIILambdaextension}
\rho^2=\frac{2\n}{p}-\frac{\Lambda}{3}p^2\,,
\qquad \zeta=\frac{1}{2\n}\,z\,,
\end{eqnarray}
and the extended $BIII$-metric takes the form
\begin{equation}\label{eq:BIIILambdaextensionmetric}
\dif s^2=4\n^2\Big[R_1(\rho)\big(-\dif t^2+\dif q^2\big)+\rho^2\dif \zeta^2
+R_2(\rho)\,\dif\rho^2\Big]\,,
\end{equation}
where  ${R_1(\rho)}$, ${R_2(\rho)}$ are evaluated by inverting (\ref{eq:BIIILambdaextension}). The allowed ranges of $p$ and the corresponding ranges of $\rho$ are summarized in Table~\ref{tbl:BIIIprange}.

\section{Global structure and physical interpretation: tachyons
in (anti-)de~Sitter spacetime} \label{sc:PhysLambda}

The metric describing both internal Regions~1 and~2 around the
tachyonic source is the $AII$-metric with $\Lambda$, namely
\begin{equation}\label{eq:AIITachLambda}
\dif s^2=\sigma^2(\dif \p^2+\sinh^2\p\,\dif\q^2)+\Big(1-\frac{2M}{\sigma}+\frac{\Lambda}{3}\,\sigma^2\Big)\dif\z^2
-\Big(1-\frac{2M}{\sigma}+\frac{\Lambda}{3}\,\sigma^2\Big)^{-1}\dif \sigma^2\,,
\end{equation}
while the external Region~3 is described by the $BI$-metric
with $\Lambda$
\begin{equation}\label{eq:BITachLambda}
\dif s^2=p^2(-\dif \tau^2+\cosh^2\tau\,\dif\q^2)+\Big(1-\frac{2M}{p}-\frac{\Lambda}{3}\,p^2\Big)\dif\z^2
+\Big(1-\frac{2M}{p}-\frac{\Lambda}{3}\,p^2\Big)^{-1}\dif p^2\,,
\end{equation}
generalizing (\ref{AIITach}) and (\ref{BITach}).

\subsection{Weak-field limit and distinct regions} \label{ssc:weakfieldLambda}
As for ${\Lambda=0}$, the key point is to consider the
weak-field limit ${M\to0}$ of (\ref{eq:AIITachLambda}) and
(\ref{eq:BITachLambda}), in which case the curvature
singularities at ${\sigma=0}$ and ${p=0}$ disappear and the
spacetimes (being then vacuum and conformally flat) can readily
be interpreted as (anti-)de~Sitter universe in which the
\emph{test tachyonic source} (located at ${\sigma=0}$ and
${p=0}$) moves with infinite speed.

\begin{figure}[ht!]
\begin{center}
\includegraphics[width=140mm]{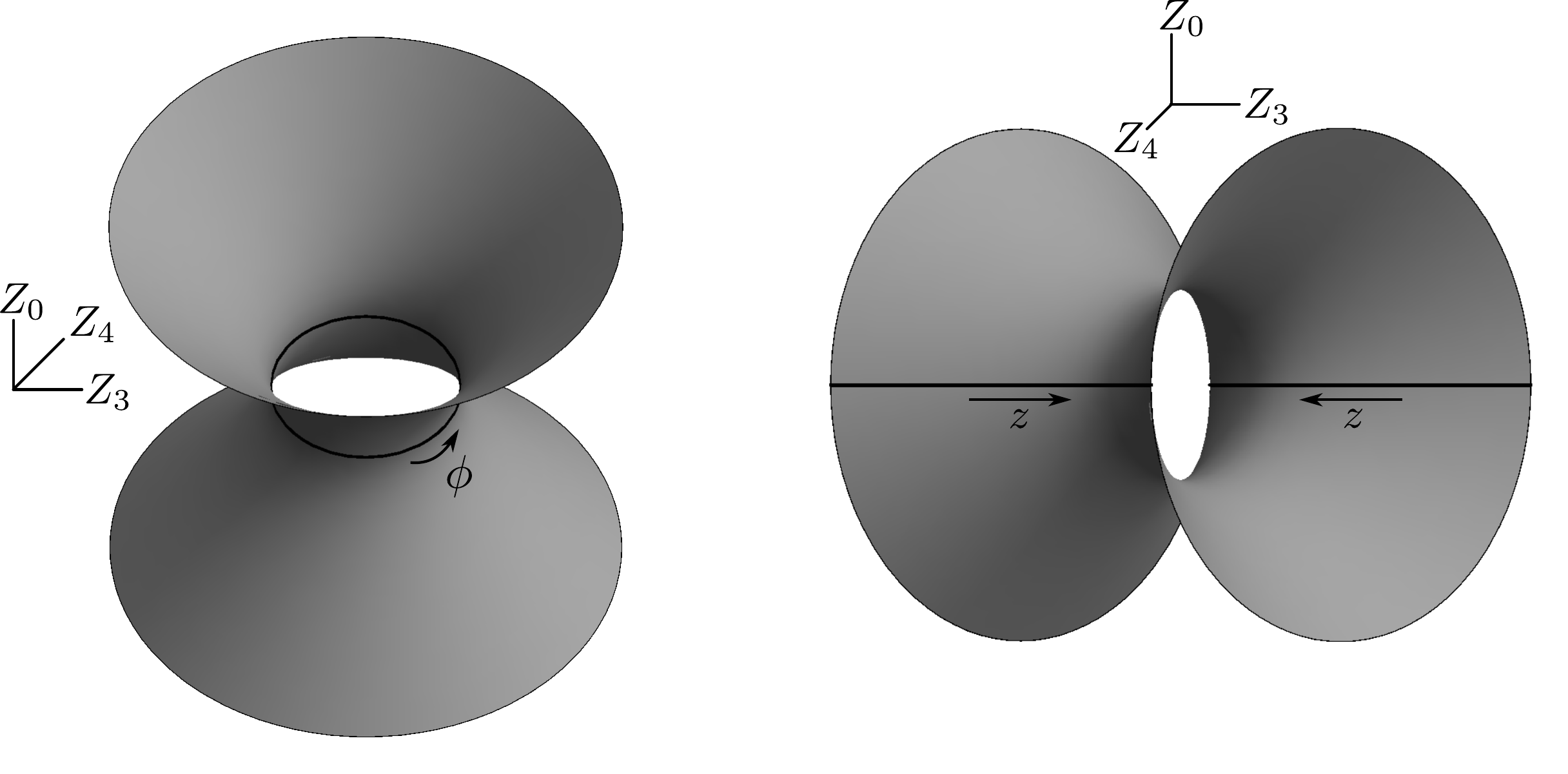}
\caption{Motion of the infinitely fast test tachyon in (anti-)de~Sitter background space.
For ${\Lambda>0}$, the tachyon runs with infinite speed in
closed circle around the neck of the de~Sitter hyperboloid (left).
For ${\Lambda<0}$, there are two tachyons running along two hyperbolic lines
on opposite sides, corresponding to the signs $\pm$ in (\ref{eq:tachtrajectory_AdS}),
of the anti-de~Sitter hyperboloid (right).}\label{img:TachdSadS}
\end{center}
\end{figure}

The trajectory of such tachyon on the hyperboloid
(\ref{eq:hyperboloids}) can be determined using the
corresponding five-dimensional parametrizations. The internal
$AII$-metric (\ref{eq:AIITachLambda}) with ${M=0}$ and
${\Lambda>0}$ is de~Sitter spacetime covered by
\begin{eqnarray}\label{eq:AIILambda+Par}
Z_{0} \rovno \pm\sigma\,\cosh\p\,,\nonumber\\
Z_{1} \rovno \sigma\,\sinh\p \,\cos \q\,,\nonumber\\
Z_{2} \rovno \sigma\,\sinh\p \,\sin \q\,,\\
Z_{3} \rovno \sqrt{\sigma^{2}+a^{2}}\,\cos \ch\,,\nonumber\\
Z_{4} \rovno \sqrt{\sigma^{2}+a^{2}}\,\sin \ch\,,\nonumber
\end{eqnarray}
which is actually (\ref{eq:AII_dS_par}) with ${p=\sigma>0}$,
${q=\pm\,\cosh \p}$ and ${t=z\equiv a\, \ch}$. The coordinate
singularity at ${\sigma=0}$ with $\p$ finite, localizing the
test tachyon, thus corresponds to the trajectory
\begin{eqnarray}\label{eq:tachtrajectory_dS}
Z_{0} \rovno 0\,,\qquad Z_{1}=0=Z_{2}\,,\nonumber\\
Z_{3} \rovno a\,\cos \ch\,,\\
Z_{4} \rovno a\,\sin \ch\,.\nonumber
\end{eqnarray}
Unlike the tachyon in Minkowski space whose trajectory is given
by the straight line (\ref{eq:tachtrajectoryMink}), which is
the $Z$-axis, this tachyon \emph{runs at infinite speed around
the neck of the de~Sitter hyperboliod}, which is the smallest
possible circle ${Z_3^2+Z_4^2=a^2}$ in such a closed universe,
see the left part of Fig.~\ref{img:TachdSadS}.

The same is true for the external $BI$-metric
(\ref{eq:BITachLambda}) with ${M=0}$ and ${\Lambda>0}$. Indeed,
the corresponding parametrization is
\begin{eqnarray}\label{eq:BILambda+Par}
Z_{0} \rovno p \,\sinh\tau\,,\nonumber\\
Z_{1} \rovno p \,\cosh\tau \,\cos\q\,,\nonumber\\
Z_{2} \rovno p \,\cosh\tau \,\sin\q\,,\\
Z_{3} \rovno \sqrt{a^2-p^2}\,\cos \ch\,,\nonumber\\
Z_{4} \rovno \sqrt{a^2-p^2}\,\sin \ch\,,\nonumber
\end{eqnarray}
equivalent to (\ref{eq:BI_adS_par}) with ${q=\sinh \tau}$,
${t=\q}$ and ${z=a\, \ch}$. Setting ${p=0}$ at finite $\tau$,
we obtain again the tachyonic trajectory
(\ref{eq:tachtrajectory_dS}).

For ${\Lambda<0}$ the tachyonic trajectory is different. The
anti-de~Sitter hyperboloid (\ref{eq:hyperboloids}) with
${\epsilon=-1}$ is parametrized in the form of the (weak-field
limit of) the $AII$-metric (\ref{eq:AIITachLambda}) as
\begin{eqnarray}\label{eq:AIILambda-Para}
Z_{0} \rovno \pm\sigma \,\cosh\p\,,\nonumber\\
Z_{1} \rovno \sigma \,\sinh\p \,\cos \q\,,\nonumber\\
Z_{2} \rovno \sigma \,\sinh\p \,\sin \q\,,\\
Z_{3} \rovno \pm\sqrt{a^2-\sigma^2}\,\sinh(\z/a)\,,\nonumber\\
Z_{4} \rovno \pm\sqrt{a^2-\sigma^2}\,\cosh(\z/a)\,,\nonumber
\end{eqnarray}
cf. (\ref{eq:AII_adS_par}). The test tachyon trajectory given
by ${\sigma=0}$ with $\p$ finite is thus located at
\begin{eqnarray}\label{eq:tachtrajectory_AdS}
Z_{0} \rovno 0\,,\qquad Z_{1}=0=Z_{2}\,,\nonumber\\
Z_{3} \rovno \pm a\,\sinh (\z/a)\,,\\
Z_{4} \rovno \pm a\,\cosh (\z/a)\,.\nonumber
\end{eqnarray}
There are thus \emph{two} tachyons \emph{moving at infinite
speed along main hyperbolic lines ${Z_4^2-Z_3^2=a^2}$  on
opposite sides of the anti-de~Sitter hyperboloid}, as
illustrated on the right part of Fig.~\ref{img:TachdSadS}.

The same result is obtained for the $BI$-metric
(\ref{eq:BITachLambda}), which for ${\Lambda<0}$ corresponds to
\begin{eqnarray}\label{eq:BILambda-Par}
Z_{0} \rovno p \,\sinh\tau\,,\nonumber\\
Z_{1} \rovno p \,\cosh\tau \,\cos\q\,,\nonumber\\
Z_{2} \rovno p \,\cosh\tau \,\sin\q\,,\\
Z_{3} \rovno \pm\sqrt{a^2+p^2}\,\sinh(\z/a)\,,\nonumber\\
Z_{4} \rovno \pm\sqrt{a^2+p^2}\,\cosh(\z/a)\,,\nonumber
\end{eqnarray}
see (\ref{eq:BI_adS_par}). For ${p=0}$ and finite $\tau$, we
recover the same tachyonic trajectory
(\ref{eq:tachtrajectory_AdS}).

\begin{figure}[b!]
\begin{center}
\includegraphics[width=150mm]{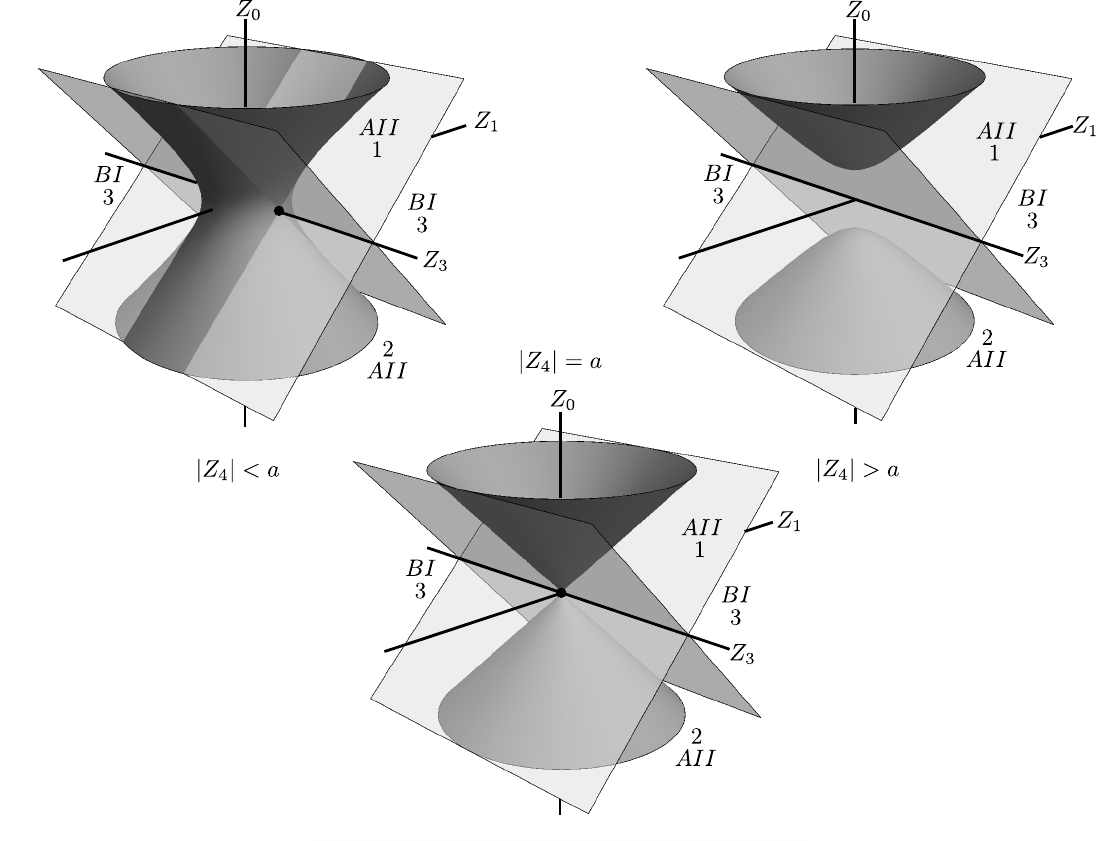}
\caption{Separation of the background de~Sitter space into Regions~1,~2 and~3,
covered by the $AII$ and $BI$-metrics, for ${Z_2=0}$ and ${|Z_4|<a}$
(upper left part), ${|Z_4|>a}$ (upper right part) and ${|Z_4|=a}$ (lower part).
For ${|Z_4|<a}$ the separation surface ${Z_1^2+Z_2^2=Z_0^2}$ cuts the de~Sitter hyperboloid,
for ${|Z_4|=a}$ it is tangent to it, and for ${|Z_4|>a}$
this surface does not intersect the hyperboloid.
Because the coordinate $Z_4$ is suppressed here, the tachyon motion is not visualized very well:
In this section, it corresponds just to two points at the intersection of the $Z_3$-axis with
the hyperboloid when ${|Z_4|<a}$, and one point when ${|Z_4|=a}$. The tachyon motion is better seen in
Fig.~\ref{img:TachdSadS} where both nontrivial coordinates $Z_3$ and $Z_4$ are visible.}
\label{img:TachBckdS}
\end{center}
\end{figure}

\subsection{Mach--Cherenkov shockwave separating the $AII$ and
$BI$-metrics}

In the weak-field limit, a pair of metrics
(\ref{eq:AIITachLambda}) and (\ref{eq:BITachLambda}) together
cover the full (anti-)de~Sitter universe. Internal Region~1 and
Region~2 are localized at ${Z_0^2>Z_1^2+Z_2^2}$ with ${Z_0>0}$
and ${Z_0<0}$, respectively. They are represented by two
$AII$-metrics (\ref{eq:AIITachLambda}). The complementary
external Region~3 represented by a single $BI$-metric
(\ref{eq:BITachLambda}) is localized at ${Z_0^2<Z_1^2+Z_2^2}$.

These regions are \emph{separated by the surface}
${Z_1^2+Z_2^2=Z_0^2}$, with $Z_3$ arbitrary such that
${Z_3^2=\epsilon(a^2-Z_4^2)}$, corresponding to the singularity
${\sigma=0,\p=\infty}$ and ${p=0, \tau=\infty}$, respectively.
This surface represents the \emph{Mach--Cherenkov shocks}, the
\emph{contracting one for} ${Z_0<0}$ given by
${\sqrt{Z_1^2+Z_2^2}=-Z_0}$ and the \emph{expanding one for}
${Z_0>0}$ given by ${\sqrt{Z_1^2+Z_2^2}=Z_0}$. As in the case
of Minkowski background, visualized in Fig.~\ref{img:TachBck},
 it is locally
 a \emph{contracting/expanding cylinder around the
superluminal tachyonic source}. However, in the ${\Lambda>0}$
case, this cylinder is ``wraped'' around the circular
trajectory (\ref{eq:tachtrajectory_dS}) in closed de~Sitter
space, so that topologically and also geometrically it is a
\emph{toroidal surface}. For ${\Lambda<0}$, instead, there are
\emph{two infinite cylinders} around
(\ref{eq:tachtrajectory_AdS}) on opposite sides of the
anti-de~Sitter hyperbolic space, see Fig.~\ref{img:TachdSadS}.

\begin{figure}[ht!]
\begin{center}
\includegraphics[width=75mm]{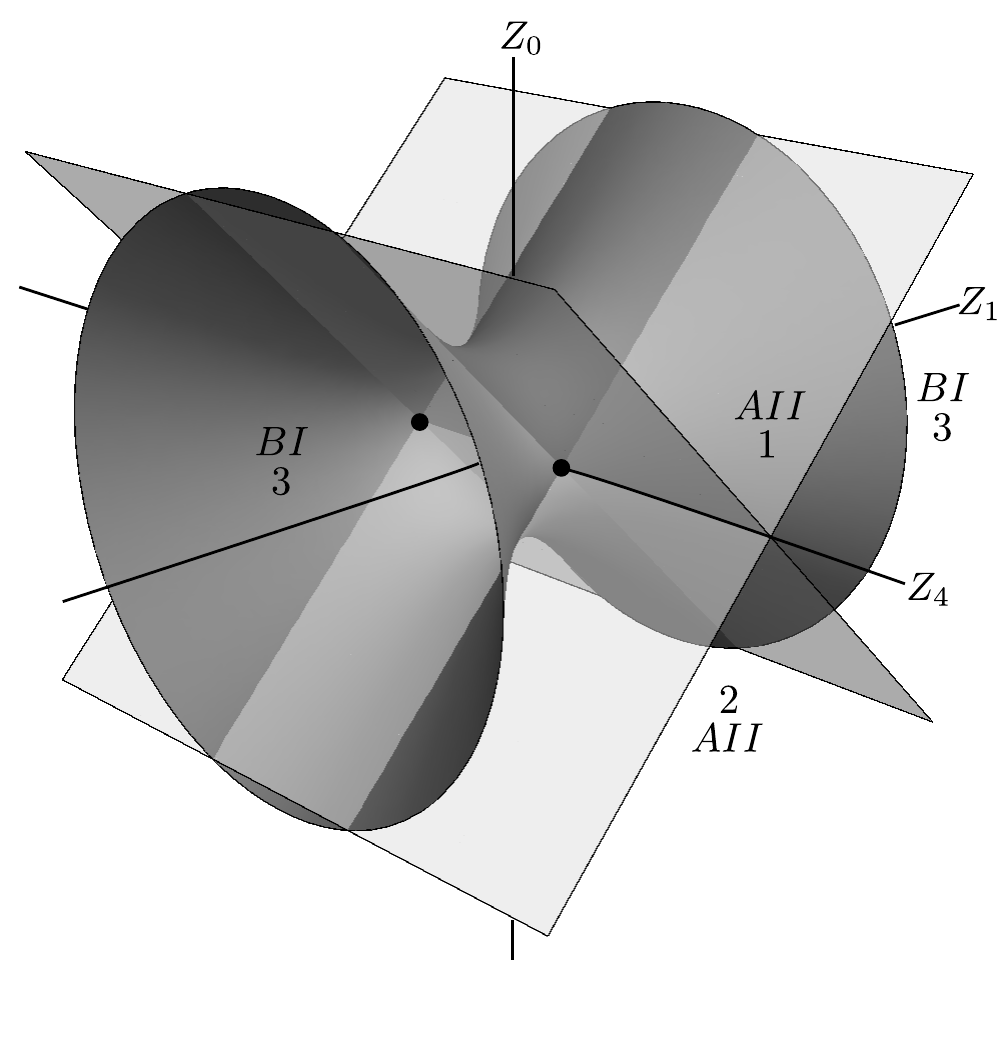}
\caption{Separation of the background anti-de~Sitter space into Regions~1,~2 and~3
for the section ${Z_2=0}$ with ${Z_3=\konst}$ Unlike for de~Sitter space, visualized
in Fig.\ref{img:TachBckdS}, this picture looks qualitatively the same for any
value of $Z_3$. Motion of the tachyon is better seen in Fig.~\ref{img:TachdSadS}.}
\label{img:TachBckAdS}
\end{center}
\end{figure}

The position of these shock surfaces, separating the $AII$ and
$BI$ regions, is shown in Fig.~\ref{img:TachBckdS} for typical
three-dimensional sections (given by three distinct values of
$Z_4$) through de~Sitter space, and in
Fig.~\ref{img:TachBckAdS} for anti-de~Sitter space.

\newpage

\subsection{Boosted metrics with ${\Lambda\neq0}$}

\begin{figure}[t!]
\begin{center}
\includegraphics[width=140mm]{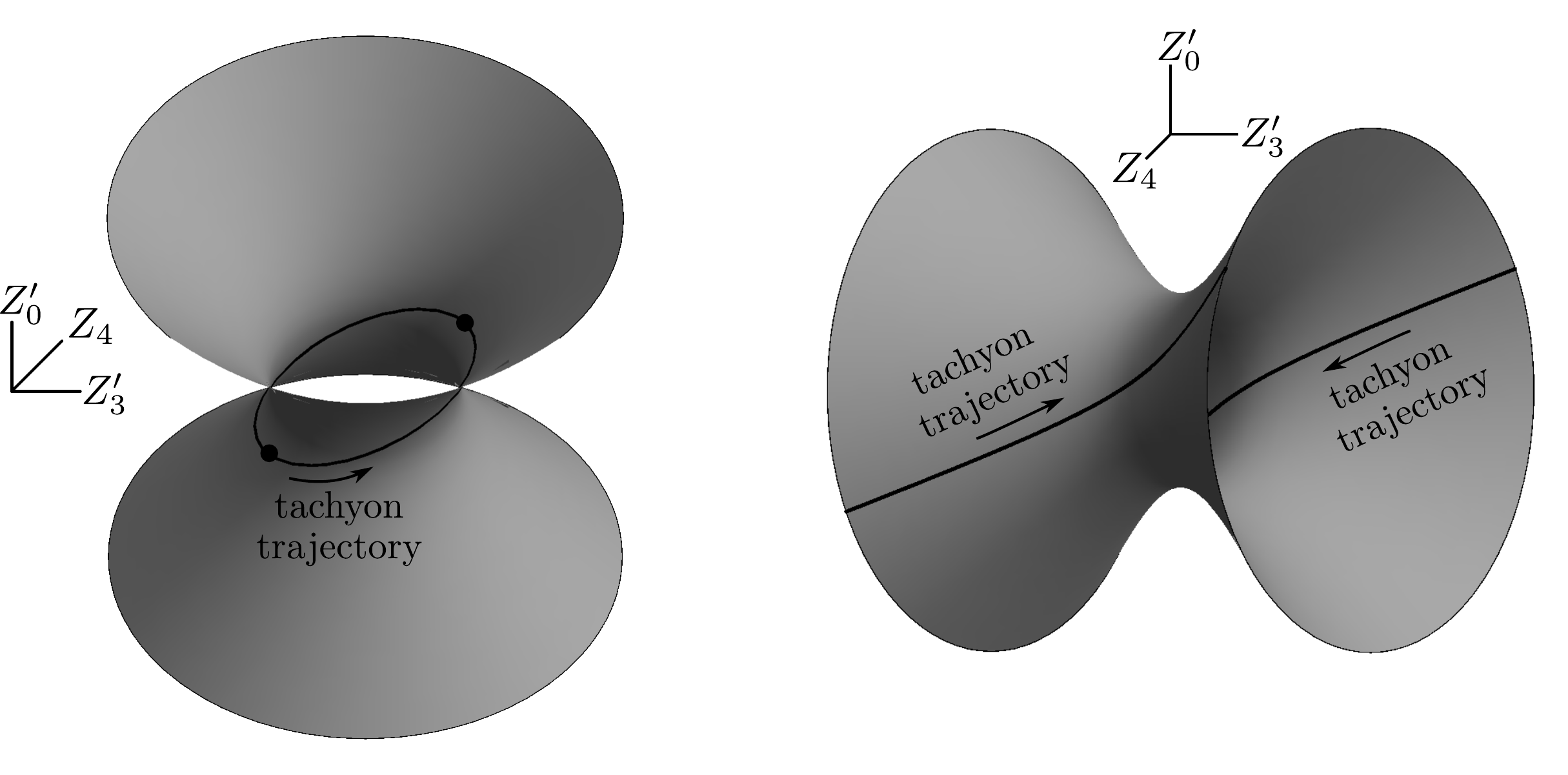}
\caption{Motion of a slowed tachyon in de~Sitter (left) and anti-de~Sitter (right)
background space in boosted coordinates. Its trajectory is given by
${Z_3'/Z_0'=v>1}$ and ${Z_1=0=Z_2}$.}\label{img:TachdSadSboost}
\end{center}
\end{figure}

Following the idea outlined in \cite{PodHru17}, it is useful to
perform a \emph{boost which slows the tachyonic source} from
infinite speed to some finite speed ${v>1}$. This enables us to
better illustrate its motion and the generated Mach--Cherenkov
conical shockwaves in (anti-)de~Sitter universe, analogously to
flat background case shown in Fig.~\ref{img:TachBckBoost}. Such
a boost must be performed in the 5-dimensional coordinates of
(\ref{eq:hyperboloids}), (\ref{eq:5D}). Choosing the spatial
$Z_3$-direction, it reads
\begin{equation}\label{eq:BoostAdSsuperlum}
 Z_0 = \frac{v\,Z_0'-Z_3'}{\sqrt{v^2-1}}\,,\qquad
 Z_3 = \frac{v\,Z_3'-Z_0'}{\sqrt{v^2-1}}\,,
\end{equation}
cf.~(\ref{eq:boost_for_AII}). The tachyon moving along
${Z_1=0=Z_2}$ at ${Z_0=0}$, see (\ref{eq:tachtrajectory_dS}),
(\ref{eq:tachtrajectory_AdS}), is then located at
${Z_3'=v\,Z_0'}$, and the shockwave surface
${Z_1^2+Z_2^2=Z_0^2}$ becomes
\begin{equation}\label{eq:CerconeAdS}
Z_1^2+Z_2^2=\frac{(v\,Z_0'-Z_3')^2}{v^2-1}.
\end{equation}
The \emph{trajectory} of thus slowed tachyon in
(anti-)de~Sitter universe is illustrated in
Fig.~\ref{img:TachdSadSboost}.

\begin{figure}[t!]
\begin{center}
\includegraphics[width=150mm]{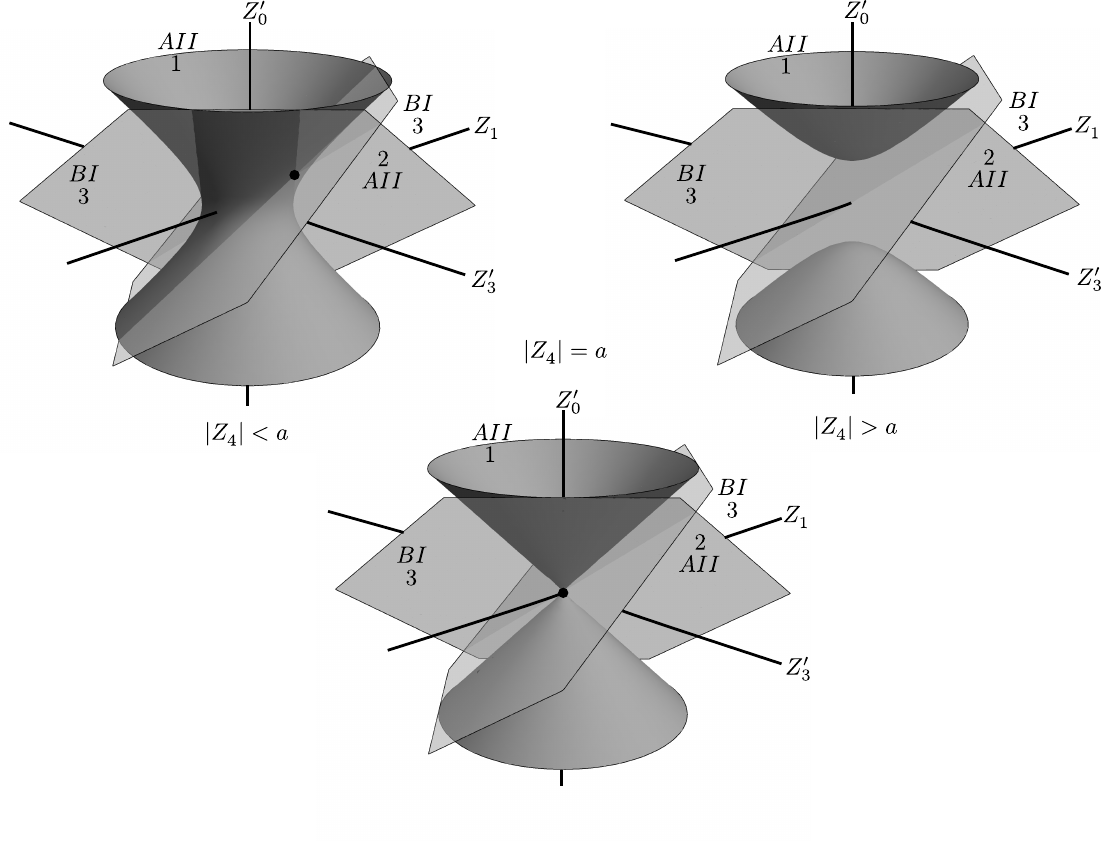}
\caption{Separation of de~Sitter universe into Regions~1,~2 and~3 in the boosted coordinates for
${|Z_4|<a}$ (upper left part), ${|Z_4|>a}$ (upper right part) and ${|Z_4|=a}$ (lower part), with  ${Z_2=0}$.
The intersections of these separation boundaries with the hyperboloid in the upper left part give the position of the
Mach--Cherenkov shock cones in de~Sitter universe. The tachyon is always located at their
joint vertex at ${Z_1=0=Z_2}$.}\label{img:TachBckdSboost}
\end{center}
\end{figure}

Such a motion in de~Sitter space admits two interpretations.
Either the tachyon moves forward in time from the initial point
(starting at ${Z_0'=-Z_{0\,\rm{max}}'}$) to the final point
(reaching it at ${Z_0'=+Z_{0\,\rm{max}}'}$), and then travels
\emph{backwards in time} to the initial point, also at speed
faster than light. Alternatively, the tachyon moves from the
initial to the final point along \emph{both trajectories}
(recall that there is not a unique geodesic between two events
in pseudo-Riemannian geometry). The specific value
${Z_{0\,\rm{max}}'}$ can be evaluated. Substituting the
conditions ${Z_1=0=Z_2}$, ${Z_3'=v\,Z_0'}$ into the boosted
form of (\ref{eq:hyperboloids}), namely
${-Z_0'^2+Z_1^2+Z_2^2+Z_3'^2+Z_4^2=a^2}$, we obtain
${Z_0'^2=(a^2-Z_4^2)/(v^2-1)}$. This explicitly determines the
position of the tachyon on the de~Sitter hyperboloid as a
function of time $Z_0'$. Extreme values of $Z_0'$ arise when
${Z_4=0}$, yielding
\begin{equation}\label{eq:Z0max}
 Z_{0\,\rm{max}}' = \frac{a}{\sqrt{v^2-1}}\,.
\end{equation}

In anti-de~Sitter space there are \emph{two tachyons}, each
moving faster than light on the opposite sides of the universe,
see the right part of Fig.~\ref{img:TachdSadSboost}.
``Unfolding'' the hyperboloid, there would be just one tachyon
oscillating between ${Z_3',Z_4=\pm\infty}$, that is ``bouncing
off'' the conformal infinity.

Visualization of Regions 1, 2 and 3, separated by the conical
shockwave surface (\ref{eq:CerconeAdS}) in the \emph{boosted
coordinates of de~Sitter space}, is shown in
Fig.~\ref{img:TachBckdSboost}. At any time $Z'_0$, this
separation boundaries localize on the hyperboloid the
Mach--Cherenkov shock, namely the \emph{rear expanding cone}
and the \emph{forward contracting cone}. Analogously as in the
situation shown in Fig.~\ref{img:TachBckBoost}, the tachyon is
always located at the intersection of these two cones at
${Z_1=0=Z_2}$.

\begin{figure}[t!]
\begin{center}
\includegraphics[width=85mm]{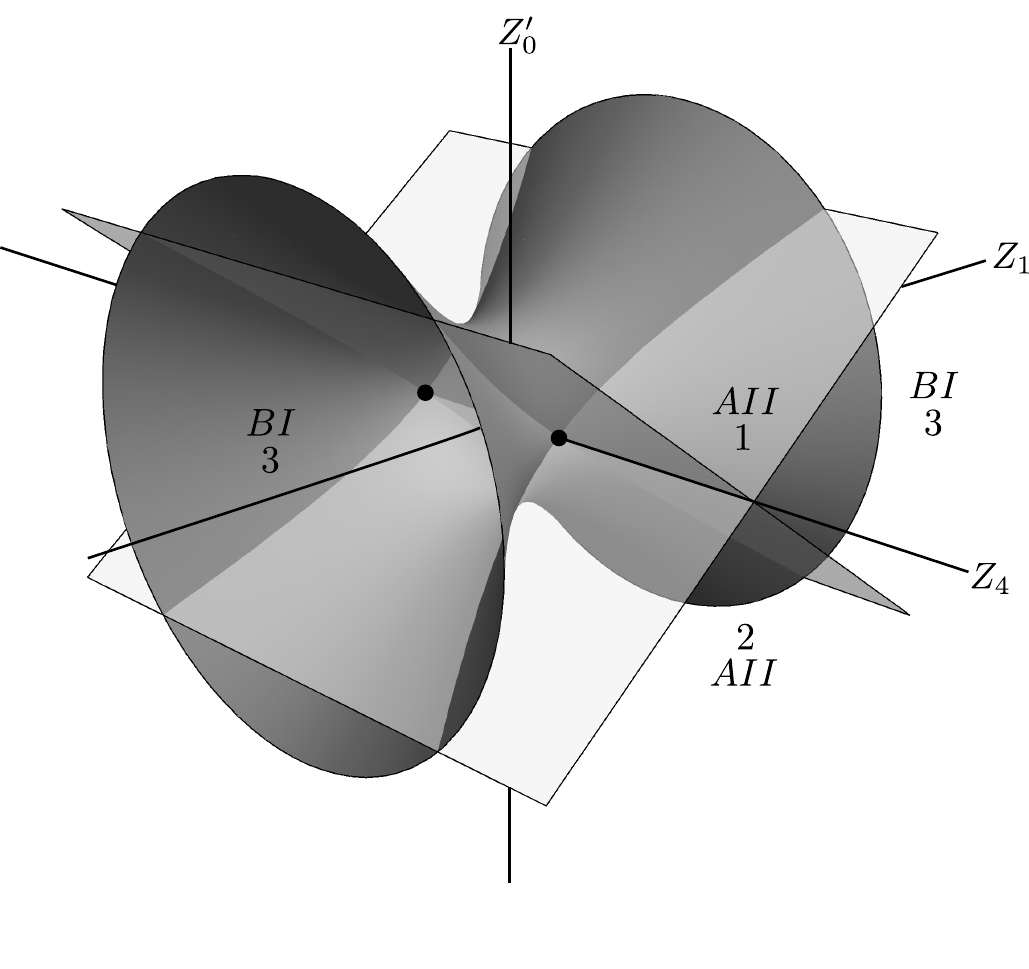}
\caption{Separation of anti-de~Sitter universe into Regions~1,~2 and~3 in the
boosted coordinates. Notice that Regions 1~and~2, covered by the $AII$-metrics,
are larger, whereas Region~3 covered by the $BI$-metric is smaller, compared to Fig.~\ref{img:TachBckAdS} (i.e.,
without the boost).}\label{img:TachBckdASboost}
\end{center}
\end{figure}

Similarly, in Fig.~\ref{img:TachBckdASboost} we visualize these
separation boundaries and the corresponding two conical shocks
in the \emph{boosted form of anti-de~Sitter space}.

\begin{figure}[t!]
\begin{center}
\includegraphics[width=150mm]{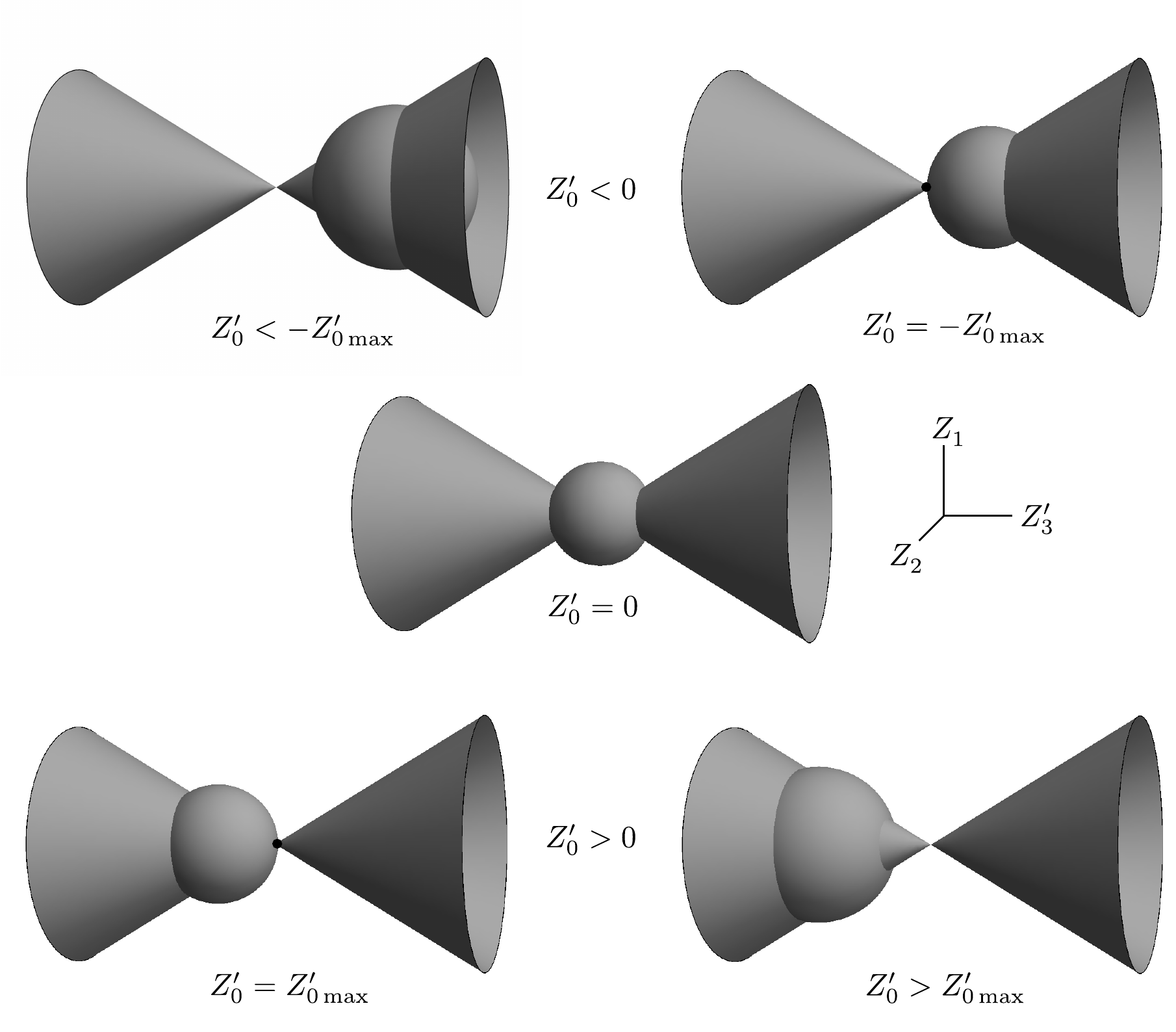}
\caption{
A time sequence visualizing de~Sitter universe --- in this section
${Z_1, Z_2, Z_3'}$ represented as a sphere --- at five
different times ${Z_0'=\,}$const. (and fixed ${Z_4<a}$). The universe
contracts, reaches its minimal size at ${Z_0'=0}$, and then re-expands.
The intersections of the sphere with the cones give the actual
position of the two Mach--Cherenkov shocks in the universe.
For ${Z_0'< -Z_{0\,\rm{max}}'}$ both shocks are contracting on the
sphere towards its Poles, while for
${Z_0'>Z_{0\,\rm{max}}'}$ they are both expanding from the Poles.
The tachyon is always located at the
joint vertex of the cones.
Analogous pictures apply to anti-de~Sitter universe expressed in
global static coordinates because the 2-spaces of constant $T$ and $r$ are also spheres,
see the metric (5.4) in \cite{GriPod09}.}\label{img:TachCherenkovdS}
\end{center}
\end{figure}

Finally, it is illustrative to visualize de~Sitter spacetime,
together with the actual position of the Mach--Cherenkov
shocks, in \emph{spatial sections} given by ${Z_1, Z_2, Z_3'}$
(with a fixed value of ${Z_4<a}$). This is done in
Fig.~\ref{img:TachCherenkovdS} for five different times
${Z_0'=\,}$const. In any such  section, the de~Sitter spatial
geometry is a 3-sphere, represented here as a 2-sphere because
one spatial dimension is suppressed in this plot. As a function
of ${Z_0'}$, this \emph{de~Sitter sphere contracts to a minimal
size at ${Z_0'=0}$ and then re-expands}. The shock surface
given by (\ref{eq:CerconeAdS}) has the form of two cones, with
the tachyon located at their joint vertex. The position of the
Mach--Cherenkov shocks in de~Sitter universe is given here by
the \emph{circular intersection of these cones with each
sphere}. There are two shocks, the forward shock and the rear
one. Recall that these shocks separate the internal Regions~1
and~2 (with the $AII$-metric) from the external Region~3
(endowed with the $BI$-metric).

The top left part of Fig.~\ref{img:TachCherenkovdS} shows the
situation in a generic time ${Z_0'< -Z_{0\,\rm{max}}'}$, in
which case both the shocks are contracting from the equator of
the spherical de~Sitter space, approaching the North Pole and
the South pole (both located at ${Z_1=0=Z_2}$), respectively.
At the special time ${Z_0'= -Z_{0\,\rm{max}}'}$, shown in the
top right part, the tachyon occurs in the North Pole of the
space when the first contracting shock reaches it, crosses it,
and starts to re-expand. The second shock on the southern
hemisphere continues to contract towards the South pole. At
${Z_0'=0}$, see the middle part of
Fig.~\ref{img:TachCherenkovdS}, the situation is fully
symmetric: The closed de~Sitter universe has a minimal radius,
both shock waves are of the same size, and are located
symmetrically with respect to the equator. For ${Z_0'>0}$, the
situation is complementary to the top part of the figure.
Bottom left part shows the tachyon located in the South Pole at
time ${Z_0'= Z_{0\,\rm{max}}'}$ when the second contracting
shock has just shrinked to zero and starts to re-expand from
the South Pole, while the first shock had been already expanding from
the North Pole. A generic situation at a time
${Z_0'>Z_{0\,\rm{max}}'}$ is shown in the bottom right part of
Fig.~\ref{img:TachCherenkovdS}, with two expanding shocks in
the expanding de~Sitter universe, both of them approaching the
equator.

\section{Conclusions}

We have presented and analyzed the classes of $A$ and
$B$-metrics with an arbitrary value of the cosmological
constant. While the famous Schwarzschild--(anti-)de~Sitter
spacetime (which is the $AI$-metric) represents the spherically
symmetric gravitational field of a static massive source, the
$AII$ and $BI$-metrics describe the field of a superluminal
source, i.e. tachyon moving along the axis of symmetry. In
fact, all these three families of metrics are related by an
appropriate boost (admitting speeds ${v>1}$).

We have studied the weak-field limit, analytic extensions, and
global structure of these spacetimes. We have demonstrated that
the full gravitational field of a tachyon in Minkowski or
(anti-)de~Sitter universe is obtained by combining a pair of
$AII$-metrics with a single $BI$-metric. The former represent
the contracting/expanding interior regions while the latter
represents an exterior region with respect to the separation
boundary which is the contracting/expanding Mach--Cherenkov
shockwave. This structure of the ``composite spacetime'',
yielding the complete gravitational field of a tachyon moving
with any superluminal speed in Minkowski, de~Sitter or
anti-de~Sitter universe, has been analyzed and visualized on
numerous pictures.

In fact, the present work is the third paper in our recent
series which we have devoted to deeper geometric, algebraic,
and physical investigation of a large family on non-expanding
Pleba\'nski--Demia\'nski space-times. This whole family,
generalizing the original $B$-metrics of \cite{EhlersKundt62},
was described and its free parameters were identified and
studied in our work \cite{PodHruGri18}. A thorough
investigation of the character of the corresponding background
coordinates for de~Sitter and anti-de~Sitter universe was
presented in \cite{PodHru17}. We hope that, together with this
third complementary paper, we have thus provided an extensive
survey and review of this simple yet interesting family of
exact solutions of Einstein's field equations.

\section*{Acknowledgements}

This work was supported by the Czech Science Foundation grant
GA\v{C}R 17-01625S. O.H. also acknowledges the support by the
Charles University Grant GAUK~196516.

\end{document}